\documentclass[11pt,twoside]{article}
\usepackage{mathptmx}
\usepackage{tabularx}
\usepackage{subcaption, arydshln}
\usepackage{geometry}
\usepackage{rotating}
\usepackage{aas_macros}
\usepackage[utf8]{inputenc}
\usepackage{fancyhdr,lastpage}
\usepackage[english]{babel}
\usepackage{enumitem,amssymb}
\usepackage{ragged2e}
\newlist{thematic}{itemize}{8}
\setlist[thematic]{label=$\square$}
\usepackage{pifont}
\usepackage{cite}
\usepackage{soul}
\usepackage{lscape}
\usepackage{multicol}
\usepackage{multirow}
\usepackage{array}
\usepackage{wrapfig}
\usepackage{graphicx}
\usepackage[range-units=brackets, range-phrase=-]{siunitx}
\usepackage{url}
\usepackage{xcolor}
\definecolor{dullpurple}{rgb}{0.431,0.188,0.534}
\definecolor{darkgreen}{rgb}{0.075,0.302,0.047}
\definecolor{dullred}{rgb}{0.706,0.208,0.192}
\usepackage[colorlinks=true, urlcolor=dullpurple, citecolor=dullred, linkcolor=darkgreen]{hyperref}
\usepackage{hyperref}
\usepackage[babel=true]{microtype}

\newlength{\enumindent}
\setlist{nolistsep}

\usepackage[font=small]{caption}
\newcolumntype{C}[1]{>{\centering\let\newline\\\arraybackslash\hspace{0pt}}m{#1}}

\DeclareSIUnit{\parsec}{pc}
\DeclareSIUnit{\Mpc}{\mega\parsec}
\DeclareSIUnit{\h}{\mathit{h}}
\DeclareSIUnit{\hPerMpc}{\h\per\Mpc}
\newcommand{\hMpc}{h {\rm Mpc}^{-1}}

\newcommand{\minisection}[1]{\noindent \textbf{#1}}

\makeatletter
\renewcommand\section{\@startsection{section}{1}{\z@}%
	{-1.5ex \@plus -0.4ex \@minus -.2ex}%
	{0.8ex \@plus.2ex \@minus .2ex}%
	{\normalfont\Large\bfseries}}
\renewcommand\subsection{\@startsection{subsection}{2}{\z@}%
	{-0.7ex\@plus -0.3ex \@minus -.2ex}%
	{0.4ex \@plus .1ex \@minus .1ex}%
	{\normalfont\large\bfseries}}
\makeatother

\newcommand{\stagetwo}{Stage~{\sc ii}}
\newcommand{\stageone}{Stage~{\sc i}}
\newcommand{\fnl}{f_{\rm NL}}

\newcommand{\BNL}{Brookhaven National Laboratory, Upton, NY~11973, USA}

\newcommand{\CERN}{Theoretical Physics Department, CERN, 1211 Geneva 23, Switzerland}

\newcommand{\CITA}{Canadian Institute for Theoretical Astrophysics, University of Toronto, Toronto, ON~M5S~3H8, Canada}

\newcommand{\IAS}{Institute for Advanced Study, Princeton, NJ~08540, USA}

\newcommand{\LBL}{Lawrence Berkeley National Laboratory, Berkeley, CA~94720, USA}

\newcommand{\McGill}{McGill University, Montreal, QC~H3A~2T8, Canada}

\newcommand{\MIT}{Massachusetts Institute of Technology, Cambridge, MA~02139, USA}

\newcommand{\PI}{Perimeter Institute, Waterloo, ON~N2L~2Y5, Canada}

\newcommand{\UBC}{University of British Columbia, Vancouver, BC V6T 1Z1, Canada}
\newcommand{\UCB}{Department of Astronomy, University of California Berkeley, Berkeley, CA~94720, USA}
\newcommand{\UCBP}{Department of Physics, University of California Berkeley, Berkeley, CA~94720, USA}

\newcommand{\UCSD}{University of California San Diego, La Jolla, CA~92093, USA}

\newcommand{\UWMadison}{Department of Physics, University of Wisconsin-Madison, Madison, WI~53706, USA}

\newcommand{\UWC}{Department of Physics \& Astronomy, University of the Western Cape, Cape Town 7535, South Africa}

\newcommand{\VSI}{Van Swinderen Institute for Particle Physics and Gravity, University of Groningen, 9747~AG~Groningen, The~Netherlands}

\newcommand{\WCA}{Waterloo Centre for Astrophysics, University of Waterloo, Waterloo, ON~N2L~3G1, Canada}

\newcommand{\Yale}{Department of Physics, Yale University, New Haven, CT~06520, USA}

\begin{document}

\pagenumbering{gobble}

\Large
\noindent Packed Ultra-wideband Mapping Array (PUMA\footnote{\url{https://www.puma.bnl.gov}}):\\
A Radio Telescope for Cosmology and Transients

\normalsize

\noindent Emanuele Castorina$^{1,2}$, 
Simon Foreman$^{3}$, 
Dionysios Karagiannis$^{4}$, 
Adrian Liu$^{5}$, 
Kiyoshi W. Masui$^{6}$, 
Pieter D. Meerburg$^{7}$, 
Laura B. Newburgh$^{8}$, 
Paul O'Connor$^{9}$, 
Andrej Obuljen$^{10}$, 
Hamsa Padmanabhan$^{11}$, 
J. Richard Shaw$^{12}$, 
An\v{z}e Slosar$^{9}$, 
Paul Stankus$^{9}$, 
Peter T. Timbie$^{13}$, 
Benjamin Wallisch$^{14,15}$, 
Martin White$^{2,16,17}$ 

\vspace*{0.3cm}

\normalsize
\noindent \textbf{RFI2: Submitted for consideration by the Astro2020 Decadal Survey Program Panel\\
Panel on Radio, Millimeter, and Submillimeter Observations from the Ground (RMS)  }

\vspace*{0.5cm}

\includegraphics[width=\linewidth]{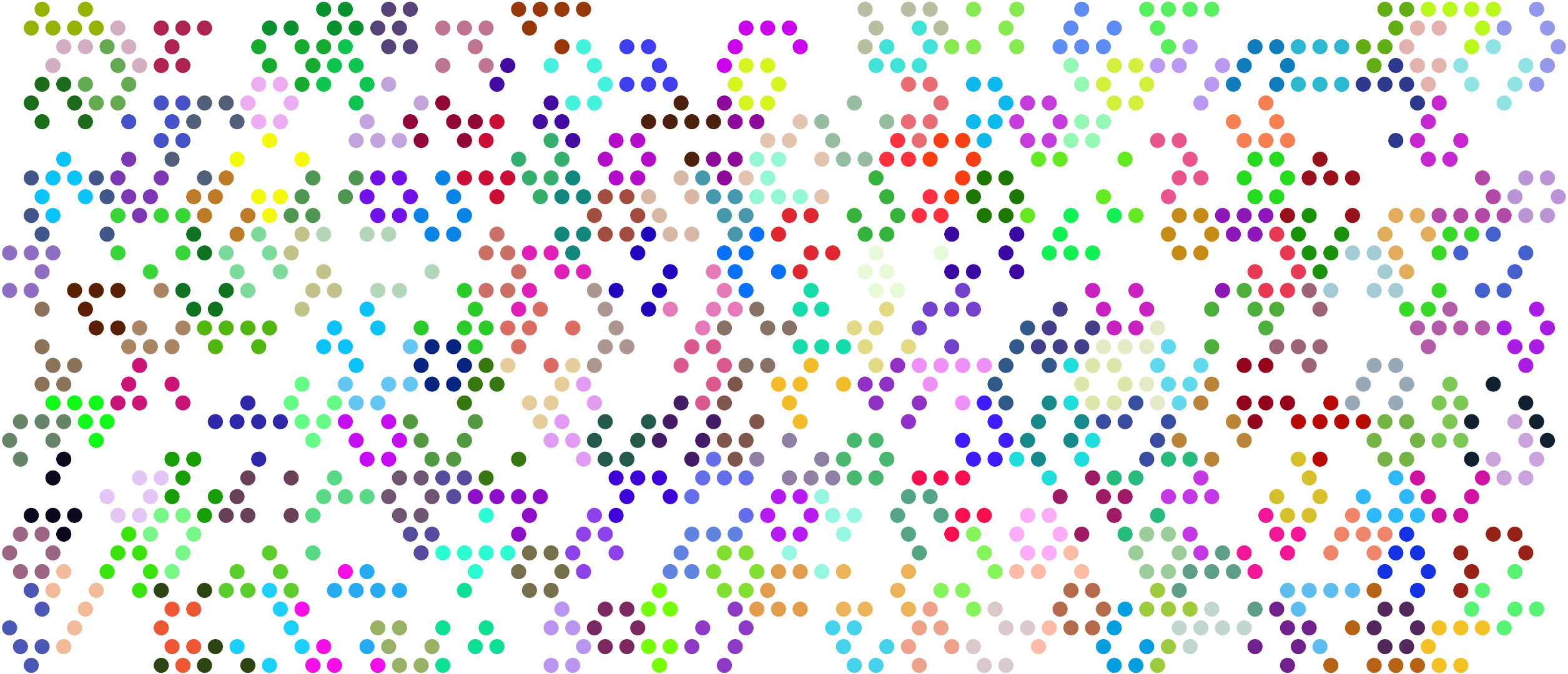}
\noindent \small \textit{ Distribution of elements in PUMA array showing a subset of 1296 elements. Elements are distributed on hexagonal lattice with 50\% occupancy rate. Clusters of 6 elements that could share the same base station with synchronized clock and a channelizer are painted in the same color.}
\vspace*{0.5cm}

{\small \noindent 
 $^{1}$ \CERN \\
$^{2}$ \UCBP \\
$^{3}$ \PI \\
$^{4}$ \UWC \\
$^{5}$ \McGill \\
$^{6}$ \MIT \\
$^{7}$ \VSI \\
$^{8}$ \Yale \\
$^{9}$ \BNL \\
$^{10}$ \WCA \\
$^{11}$ \CITA \\
$^{12}$ \UBC \\
$^{13}$ \UWMadison \\
$^{14}$ \IAS \\
$^{15}$ \UCSD \\
$^{16}$ \UCB \\
$^{17}$ \LBL \\
}

\vspace*{8cm}
The PUMA collaboration submitted the PUMA APC submission to the Decadal Survey on Astronomy and Astrophysics 2020 in July 2019. The Panel on Radio, Millimeter and Submillimeter Observations from the Ground of Astro2020 sent the collaboration a Request for Information, response to which was submitted in December 2019. Upon reviewing the submitted material, the panel had further questions that the collaboration responded in a teleconference in January 2020. This version is a lightly edited version of the document submitted in December 2019 with additional information to reflect the panel's subsequent questions. In particular:
\vspace*{0.3cm}
\begin{itemize}
\item FRB and pulsar parts of Section \ref{sec:main-science-goals} have been updated;

\item Discussion of possible funding models in Section \ref{sec:programmatic-issues} has been expanded;
  
\item Section \ref{sec:major-risks-devel} has been updated with a discussion about FFT correlation risk mitigation;

\item Cost section has been reduced in detail and moved into the Appendix \ref{app:cost}. We provide a link at which the latest costing model can be accessed. 
\end{itemize}
\newpage

\pagebreak
\pagenumbering{arabic}

\setlength{\parskip}{3pt}
\setlength{\parindent}{18pt}

\newenvironment{orangerate}{\color{orange}\begin{enumerate}}{\end{enumerate}}

\begin{centering}
\section*{Executive Summary}
\end{centering}

\subsection*{Science Objectives \& Technical Implementation}

In the next decade, observational facilities across the world will survey the Universe with exquisite precision at redshift $z<2$, but this leaves much of the cosmic volume post reionization and before the epoch of accelerated expansion uncharted. The volume is amenable to intensity mapping observations using the neutral Hydrogen 21\,cm line in a cost-effective manner, especially at redshifts beyond $z=2$, where the low bias of the distribution of neutral Hydrogen and weak non-linearity of clustering provide cosmological information to relatively small scales.

This gap led to the Packed Ultrawideband Mapping Array (PUMA) proposal, which is a large radio transit interferometric telescope. PUMA is designed to explore six basic science drivers across three different areas of physics:

\noindent \minisection{Physics of Dark Energy:}
\begin{itemize}
\item[{A.}] \textbf{Expansion History of the Universe} at sub-percent precision covering  two thirds of its evolution, from redshifts $z=6$ to $z=0.3$, including the puzzling transition from complete matter domination to dark energy domination;
\item[{B.}] \textbf{Growth of cosmic structure} using the same instrument over the same redshift range as for expansion history;
\end{itemize}

\noindent \minisection{Physics of Cosmic Inflation:}
\begin{itemize}
  \item[C.] \textbf{Primordial Non-Gaussianity} at sensitivities that significantly improve upon Cosmic Microwave Background (CMB)  measurements and are especially competitive in the equilateral and orthogonal shapes which are difficult to measure with lower redshift galaxy surveys;
  \item[D.] \textbf{Relic Inflationary Features} in the power spectrum that are a prediction of some string-inspired models;
\end{itemize}

\noindent \minisection{Time-domain Astrophysics:}
\begin{itemize}
\item[E.] \textbf{Discovery of Fast Radio Bursts (FRBs)} that will exceed all the currently proposed telescopes in the total number of observed events;
\item[F.] \textbf{Pulsar Monitoring} of a large fraction of such objects discovered by the Square Kilometer Array (SKA) telescope.
\end{itemize}

The conceptual design presented here as PUMA-32K, is aimed at a single instrument targeting these science objectives.  In brief, it consists of a five-year survey employing an array of 32,000 six-meter parabolic dishes, steerable in declination, with 0.2 - 1.1GHz bandwidth receivers located on the sites of a hexagonal close-packed lattice with 50\% occupancy. Details of the antenna and signal chain design, site considerations, calibration requirements, and data management are presented below. These parameters of the reference design are subject to modification depending on R\&D progress and funding considerations.
A smaller, 5,000-dish version of the experiment (PUMA-5K) with reduced scientific reach is also described for comparison.

\begin{figure}
  \begin{center}
  \includegraphics[width=0.8\linewidth,trim=12 45 12 12]{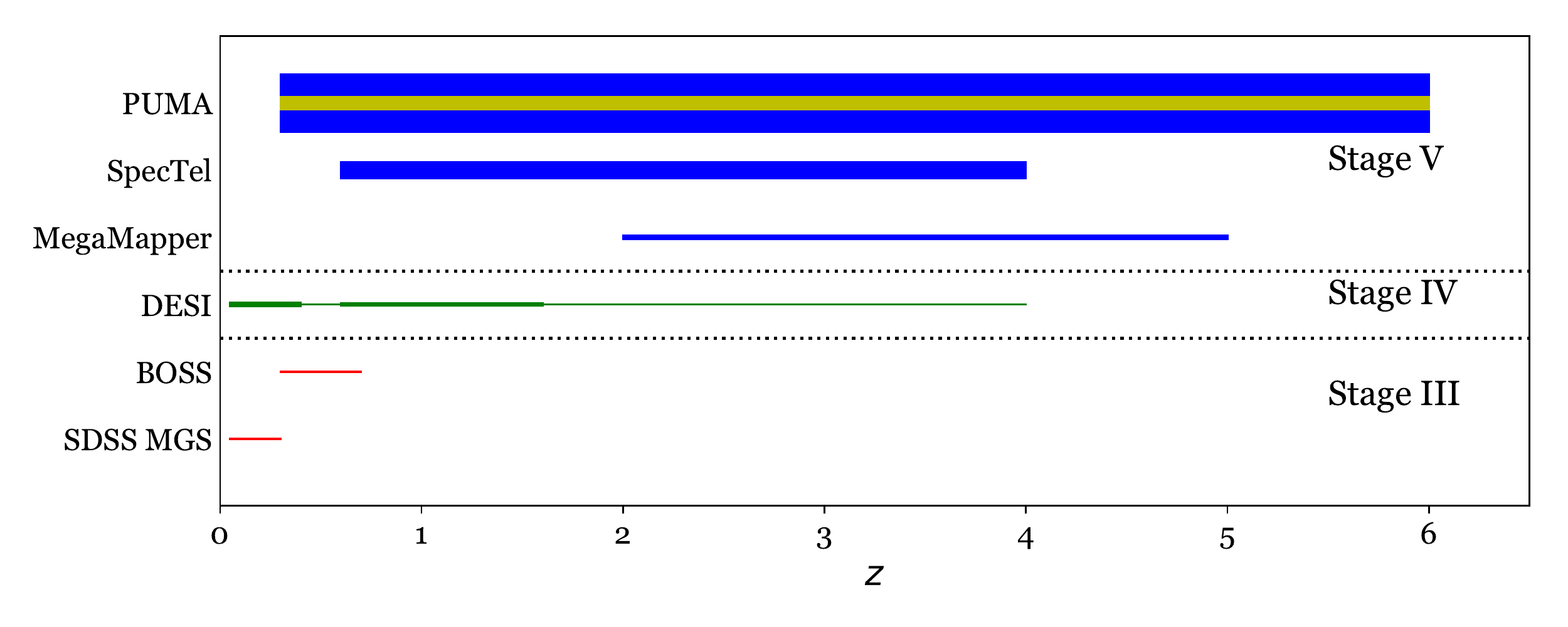}
\end{center}
  \captionsetup{font=footnotesize}
  \caption{Comparison of Cosmic Surveys sorted by Department of Energy
    Stage, including completed (Stage III), upcoming (Stage IV) and
    proposed (Stage V surveys). For each survey, the $x$-axis
    represents the approximate span in redshift, while the width of
    the line is such that the total area scales as the effective
    number of galaxies.  We assumed 100 and 480 million successful
    redshifts and for MegaMapper and SpecTel respectively.  For PUMA,
    the effective number of sources was calculated at wavenumber
    $k=\SI{0.2}{\hPerMpc}$, taking into account both the shot noise of
    the underlying galaxy population and the receiver noise after
    survey completion. PUMA numbers evaluated at $k=\SI{0.5}{\hPerMpc}$
    are $\sim 20\%$ lower. The yellow and blue bands correspond to
    PUMA-5K and PUMA-32K. While such simplistic comparison necessarily
    hides numerous caveats, it nevertheless captures the basic
    statistical reach of each experiment.}\label{fig:surveycomp}
\end{figure}

In its full configuration, PUMA's total noise will be equivalent to the sampling (Poisson) noise from a spectroscopic survey of 2.9 billion galaxies, over a redshift range largely inaccessible to optical instruments (see Fig.~\ref{fig:surveycomp} for visual comparison). Cross-correlations with optical and CMB surveys will dramatically improve our understanding of both late- and early-time cosmic acceleration, and PUMA's rich time-domain data set will catalog events in the transient radio sky, potentially in live ``multi-messenger" coincidence with other observatories operating in the 2030s.

\subsection*{Technology Maturity \& Development Roadmap}
\label{sec:techMaturity}

Unlike optical and CMB surveys, which are mature and now planning 3rd and 4th generation experiments, Hydrogen intensity mapping is a relatively new technique . There are several proposed and currently operating experiments that will use the Hydrogen intensity mapping (IM) technique at lower redshifts ($z\leq2$). These include CHIME\cite{CHIME}, BINGO\cite{wuensche2019bingo}, Tianlai\cite{wu2016tianlai}, HIRAX\cite{HIRAX}, and CHORD\cite{Vanderlinde:2019tjt}.

For a large-N interferometric array such as PUMA, the key hardware technologies include
\begin{itemize}
  \item{mass-produced, high-quality, cost-optimized optics;}
  \item{wideband, stable, low-noise analog signal chains;}
  \item{``software defined radio'' electronics to perform digitization, filtering, and temporal Fourier transform (channelization) over $\sim 1\,\mathrm{GHz}$ bandwidth;}
  \item{massively parallel digital signal processing to perform complex correlation and integration operations, extracting sky maps from raw data and reducing data volume by 5 - 6 orders of magnitude in real time;}
  \item{a means to distribute precision timing across the array to allow sub-picosecond synchronization of all receivers.}
\end{itemize}

Some of these elements have been demonstrated, albeit at smaller scale, in the current generation of IM experiments or in other applications. For PUMA, the needed developments will center around cost efficiency and quality control in large-scale manufacture. Fortunately, the electronics needed for PUMA can take direct advantage of the rapid advances in semiconductors driven by the explosive growth in wireless communications, artificial intelligence/machine learning, and big-data analytics. Based on present trends continuing, we are confident that hardware meeting PUMA's performance and cost goals will be available in the anticipated construction time frame.

For successful integration of existing technologies into a functioning experiment, more development is needed and for that we plan a 4--5 year R\&D effort by PUMA's scientific collaboration, supplemented by a series of hardware test beds. We divide this effort into four prongs:

\begin{enumerate}
\item \textbf{Technology Development in the Laboratory.} This prong includes development of both array elements, such as dishes design and construction, supporting electronics including precise clock distribution and the hardware that enables calibration.
  
\item \textbf{Computing, Software \& Pipelines.}  This prong includes development of software to enable generation of timestream data that incorporates the IM and foreground signals, antenna and signal chain non-idealities and noise, antenna-to-antenna variations, and synthetic calibration data as necessary to validate calibration and data reduction methodologies. The second aspect includes extensive electromagnetic simulators to model and predict not just beam properties, but also  coupling between array elements in detail will allow us to optimize design of those;
  
\item \textbf{Real-Time Signal Processing \& Calibration.} Both FFT-correlation and real-time calibration are necessary at the scale of PUMA, but have not been demonstrate at the required precision. This prong is focused on developing and validating  new methods and algorithms that will enable PUMA and inform hardware requirements. 
  
\item \textbf{Path-finder array.} The insight gained from the previous three prongs will be tested, to the extent possible, with smaller prototype arrays comprised of up to few tens of dishes.
\end{enumerate}

The tools developed in this R\&D phase  will allow us to follow a systems engineering process by which we will maintain and track error and noise budgets throughout the Final Design phase.

\subsection*{Areas where data for this RFI are not available.}

At this stage of the PUMA project, we are unable to give detailed answers on Facilities, Operations and Observational Strategy, although a notional decision process and elements of those that exist are discussed in the forthcoming section.

\newpage
\tableofcontents

\newpage
\section{Introduction}

This document is written in response to the National Academy of Science Decadal Survey on Astronomy \& Astrophysics (Decadal Survey) request for information by the Panel on Radio, Millimeter, and Submillimeter Observations from the Ground and responds to the questions posed in the Early-Stage Ground Concepts Questionnaire.

The PUMA concept evolved from a two year exploratory effort by the 21\,cm subgroup of the Department of Energy Cosmic Visions Dark Energy working group. Following a general call for a \stagetwo\ 21\,cm experiment \cite{Stage2WhitePaper}, we evolved the concept into a concrete proposal that culminated in the submission of an APC project to the Decadal Survey \cite{PUMAWhitePaper}.

At the time of writing the design of the PUMA experiment has been advanced to a conceptual level sufficient to allow us to forecast science results and generate an informed cost estimate (see Appendix \ref{app:cost}). A proto-collaboration is in place, whose membership includes participants in all current IM experiments, and their expert judgements have informed the PUMA conceptual design. However, no external or agency-led reviews have taken place thus far.

\section{Science}

While an instrument as powerful as PUMA can be expected to produce exciting science in areas we have not considered in detail, the facility and design are motivated by six main science drivers as presented in the Excutive Summary: \textbf{(A)} measuring the cosmic expansion history, \textbf{(B)} constraining the growth of large-scale structure and testing General Relativity, \textbf{(C)} looking for non-Gaussianity in the primordial fluctuations, \textbf{(D)} searching for features in the primordial power spectrum, \textbf{(E)} discovering FRBs and \textbf{(F)} studying the astrophysics of pulsars by monitoring many of the pulsars discovered by SKA \cite{Stage2WhitePaper}.  The first four of these objectives are enabled by a large area, high sensitivity but modest resolution map of the emission from neutral Hydrogen at $z<6$ -- known as an intensity mapping (IM) survey.  The remaining two are enabled by PUMA's continuous monitoring of the low frequency radio sky.  A number of other science goals which would be enabled by PUMA, but which did not inform the design, are discussed in \cite{Stage2WhitePaper}.  These include improved constraints on parameters of our cosmological world model, better measurements of the energy density of light relic particles from the early Universe, improved constraints on the equation of state of dark energy even in cosmologies with free neutrino mass, gravitational lensing, measures of the large-scale velocity field, multi-messenger probes and time domain surveys.

\subsection{Main Science Goals and Forecasts}
\label{sec:main-science-goals}
We will now review the six main science goals and discuss the forecasted sensitivity and how it compares to current and upcoming experiments. Much of this material is described in a larger context in \cite{Stage2WhitePaper}. We give forecasts for both PUMA-32K and the smaller version, PUMA-5K. In terms of the noise levels, the latter is roughly comparable to a spectroscopic survey of $\sim 5\times 10^8$ galaxies, and the former $\sim 2.5 \times 10^9$ galaxies. In \cite{PUMAWhitePaper} we referred to these two configurations as PUMA and Petite PUMA; some forecasted errors have changed a little from our previous papers as our forecasting codes evolve, but none of them substantially.

\begin{figure}
  \centering
  \begin{tabular}{cc}
\hspace*{-1cm}
    \includegraphics[width=0.6\linewidth,trim=0 20 0 0]{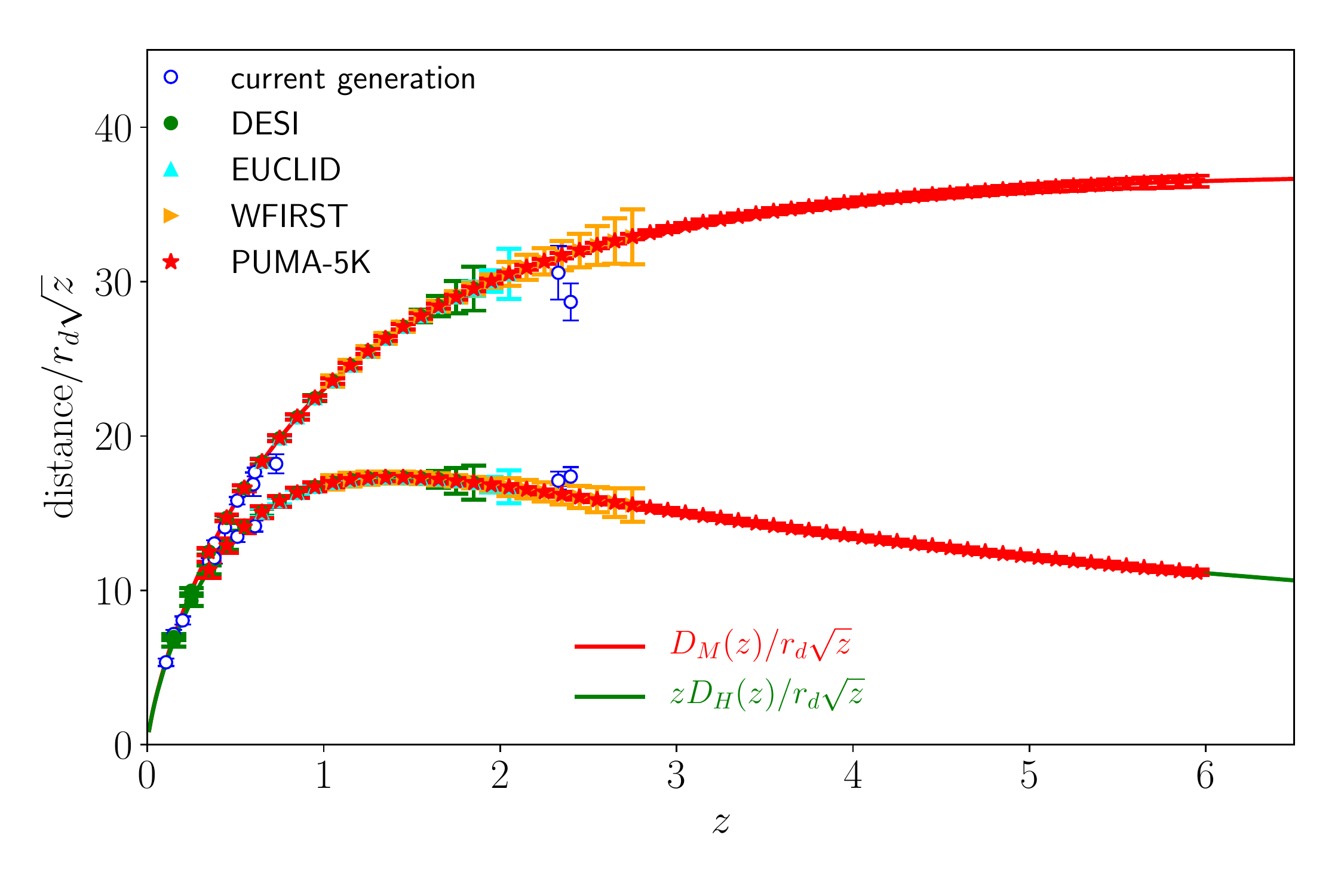} &
    \begin{minipage}{7cm}
    \vspace*{-5cm}
    \small
    \begin{tabular}{c|cc}
        redshift & PUMA-5K  & PUMA-32K \\
        \hline
1.0 & 0.7 / 1.7 (0.8 / 1.8) & 0.7 / 1.6 (0.8 / 1.7)\\
2.0 & 0.5 / 1.0 (0.8 / 1.1) & 0.4 / 0.9 (0.6 / 1.0)\\
3.0 & 0.5 / 0.9 (1.1 / 1.1) & 0.4 / 0.7 (0.8 / 0.9)\\
4.0 & 0.6 / 1.0 (2.0 / 1.3) & 0.4 / 0.7 (1.3 / 0.9)\\
5.0 & 0.8 / 1.2 (4.2 / 1.8) & 0.4 / 0.7 (2.4 / 1.1)\\
6.0 & 1.0 / 1.5 (9.7 / 2.7) & 0.4 / 0.7 (4.6 / 1.3)\\
    \end{tabular}
    \vspace*{0.1cm}

    Relative Error in percent on
    transverse / radial BAO in $\Delta z=0.1$ radial bins without reconstruction. Numbers in brackets correspond to pessimistic assumption about foreground modeling. \\
    \
    \end{minipage}\\
    \end{tabular}

    \vspace*{0.2cm}
    \caption{Constraints on the distance-redshift relation(s)
      achievable with the BAO technique for some current and up-coming
      experiments (from \cite{Stage2WhitePaper}; adapted from Fig.~1 of
      \cite{2015PhRvD..92l3516A}). Lines from top to bottom correspond
      to transverse and radial BAO for the best-fit Planck
      $\Lambda$CDM model.  The $21\,$cm errors are for the optimistic
      foreground case but with no reconstruction -- simulations
      indicate that this is conservative \cite{Modi19b} -- as described in
      the text.
      The Table on the right gives the numerical values in percent on
      the BAO precision as a function of redshift for PUMA-5K and PUMA-32K
      for transverse and radial BAO and includes ``pessimistic'' foreground
      numbers in parentheses.
      For BAO alone, the larger version does not perform significantly
      better since the measurement becomes sample variance limited already
      with PUMA-5K over much of the redshift range.}
\label{fig:bao}
\end{figure}

For the first goal \textbf{(A)}, PUMA will make use of the baryon acoustic oscillation method \cite{2013PhR...530...87W} to measure geometric distances (calibrated from the well-understood physics of the early Universe) as a function of redshift.  While existing surveys will cover the $z<2$ Universe at high precision (with weaker measurements to $z\simeq 3$, thus covering only a few percent of the observable comoving volume, e.g.\ \cite{loeb2008}) PUMA will enable nearly sample-variance limited BAO measurements all the way to $z\sim 6$ and complete the challenge of characterizing the expansion history across the cosmic ages.  The impact PUMA would have on this field can be clearly seen in Fig.~\ref{fig:bao}, bringing us a large way towards a complete measurement this fundamental property of the Universe and characterization of its (Friedmann) metric. This measurement will probe universe before the onset of dark energy and when the potential effects of curvature are largest across the cosmic history.  These forecasts are quite robust as BAO physics is well understood and the path from measurements to constraints is well developed.  The major uncertainties are the overall brightness of the $21\,$cm signal at higher redshift and the amount of the $k_\perp-k_\parallel$ plane rendered unusable by foregrounds.  In Fig.~\ref{fig:bao} we have made an ``optimistic'' foreground assumption that foreground power is largely broadband and does not affect the BAO peak.  The table also gives a more pessimistic case where the foreground wedge extends to three times the primary beam (see \cite{Stage2WhitePaper} for discussion of foreground avoidance vs.~cleaning and the impact of the foreground wedge).  In both cases we assume no gain from reconstruction, which is a pessimistic assumption \cite{Modi19b}.

\begin{figure}
  \centering
  \begin{tabular}{cc}
    \hspace*{-2cm}
    \includegraphics[width=0.75\linewidth]{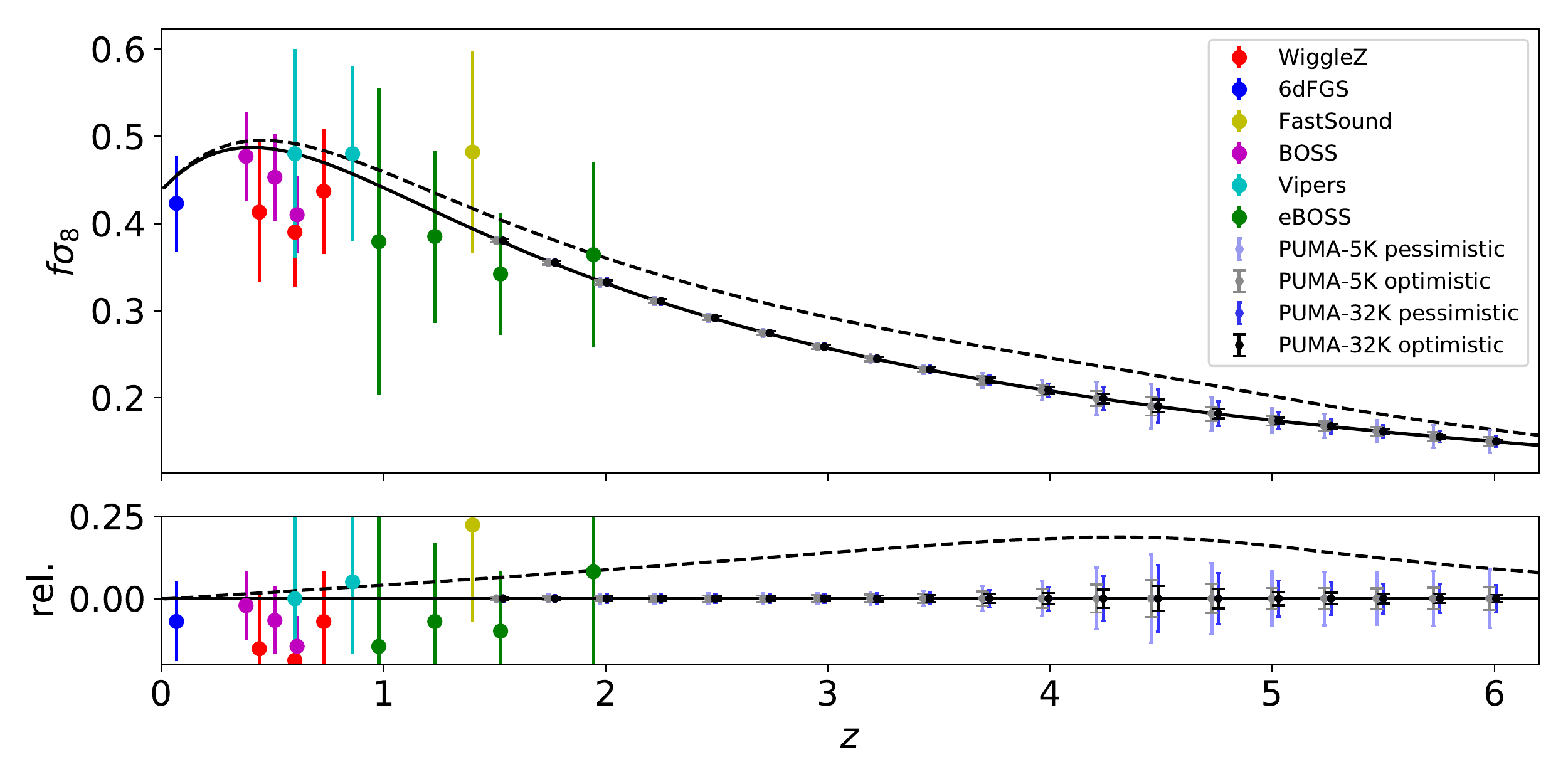}&
    \begin{minipage}{0.25\linewidth}
    \vspace*{-5.5cm}
    \small
    \begin{tabular}{c|cc}
        $z$ & PUMA-5K  & PUMA-32K \\
        \hline
    1.5 & 0.4 (0.7) & 0.3 (0.6) \\
    2.0 & 0.5 (1.0) & 0.5 (0.9) \\
    2.5 & 0.7 (1.1) & 0.6 (1.0) \\
    3.0 & 0.7 (1.1) & 0.6 (0.9) \\
    3.5 & 1.0 (1.7) & 0.7 (1.3) \\
    4.0 & 2.0 (3.8) & 1.2 (2.5) \\
    4.5 & 4.0 (9.5) & 2.7 (7.1) \\
    5.0 & 2.3 (5.8) & 1.4 (3.9) \\
    5.5 & 2.2 (5.6) & 1.0 (3.2) \\
    6.0 & 2.5 (6.3) & 0.8 (3.0)
      \end{tabular}
    \vspace*{0.1cm}
    Relative error, in percent, on $f\sigma_8$ in $\Delta z=0.5$ bins.
    \end{minipage}\\
    \end{tabular}
  \caption{Forecasts constraints on the rate-of-growth of large-scale structure ($f\sigma_8$) from $21\,$cm observations, together with a compendium of current constraints.  Lines are theoretical models: $\Lambda$CDM is plotted with a solid line while dashed is a modified gravity model (described in \cite{Stage2WhitePaper}) with vanishing effects at high redshift and an expansion history equal to that of $\Lambda$CDM. The table on the right gives the same information in tabular form, with errors assuming pessimistic foregrounds in parentheses.}
\label{fig:fsig8mg}
\end{figure}

Goal \textbf{(B)} relies upon measurement of the shape of the HI clustering signal into the quasi-linear regime \cite{2013PhR...530...87W,1902.07147}.  Since the growth of large-scale structure in the Universe is a competition between expansion (pulling things apart) and gravity (pulling things together) a measurement of the growth rate over the same redshift range as we have measurements of the expansion provides a sensitive test of General Relativity and measurement of the constituents of the Universe.  PUMA will take this measurement to a new level (Fig.~\ref{fig:fsig8mg}) in a previously unexplored regime.
While BAO measurements \textbf{(A)} rely primarily on well-localized features in the power spectrum and do not represent a major challenge in terms of measurement, the measurement of the redshift-space HI power spectrum \textbf{(B)} is more demanding (see below).  Fig.~\ref{fig:fsig8mg} compares two different assumptions about the impact of foregrounds on the useable modes in the $k_\perp-k_\parallel$ plane. The optimistic assumption assumes modes are lost in the ``primary beam wedge'' \cite{Stage2WhitePaper}, while the ``pessimistic'' assumption extends the number of lost modes to three times the primary beam.

The same measurements that allow us to test General Relativity also allow us to put tight constraints on the constituents of the Universe.  Table \ref{tab:params} shows one example, comparing constraints on the mass of the relic light neutrinos between upcoming experiments~\cite{Dvorkin:2019jgs} or constraints on additional light degrees of freedom (parameterized by $N_{\rm eff}$)~\cite{Green:2019glg}.

\begin{table}
  \centering
  \begin{tabular}{|c|c|c|c|c|c|c|}
  \hline
\multirow{3}{*}{Parameter}  & LSST & \multirow{3}{*}{CMB S4} & PUMA & LSST~+~DESI & CMB-S4 & All experiments\\
  & + DESI &  & + Planck & + PUMA & + PUMA & combined \\
  & + Planck &  & & + Planck & & \\
\hline
$\sum m_\nu$ [meV] & 38 & 59 & 31~/~27 & 25~/~22 & 24~/~21 & 15~/~14 \\
$\sum m_\nu$ + $\tau$ prior & --- & 15 & --- & --- & 14~/~13 & 10.4~/~10.2 \\
$\sum m_\nu$ (free $w$) & 50 & --- & 33~/~29 & 26~/~23 & --- & --- \\
\hline
$N_{\rm eff}$ & 0.050 & 0.026 & 0.043~/~0.037 & 0.033~/~0.030 & 0.014~/~0.013 & 0.012~/~0.011 \\
\hline
$w$ (free $\sum m_\nu$) & 0.017 & --- & 0.006~/~0.005 & 0.005~/~0.004 & --- & --- \\
\hline
\end{tabular}
\caption{\label{tab:params} Combination of parameter forecasts for a compendium of future experiments (from \cite{Stage2WhitePaper}). We tabulate $1\sigma$ forecasted uncertainties.  All combinations include a Planck 2015 CMB prior and are for a $\Lambda$CDM cosmology unless stated otherwise. For combinations involving $21\,$cm we state both the pessimistic and optimistic foreground removal cases respectively, separated by a slash.}
\end{table}

\begin{figure}
  \centering
  \begin{tabular}{cc}
    \hspace*{-2cm} 
  \includegraphics[width=0.4\linewidth]{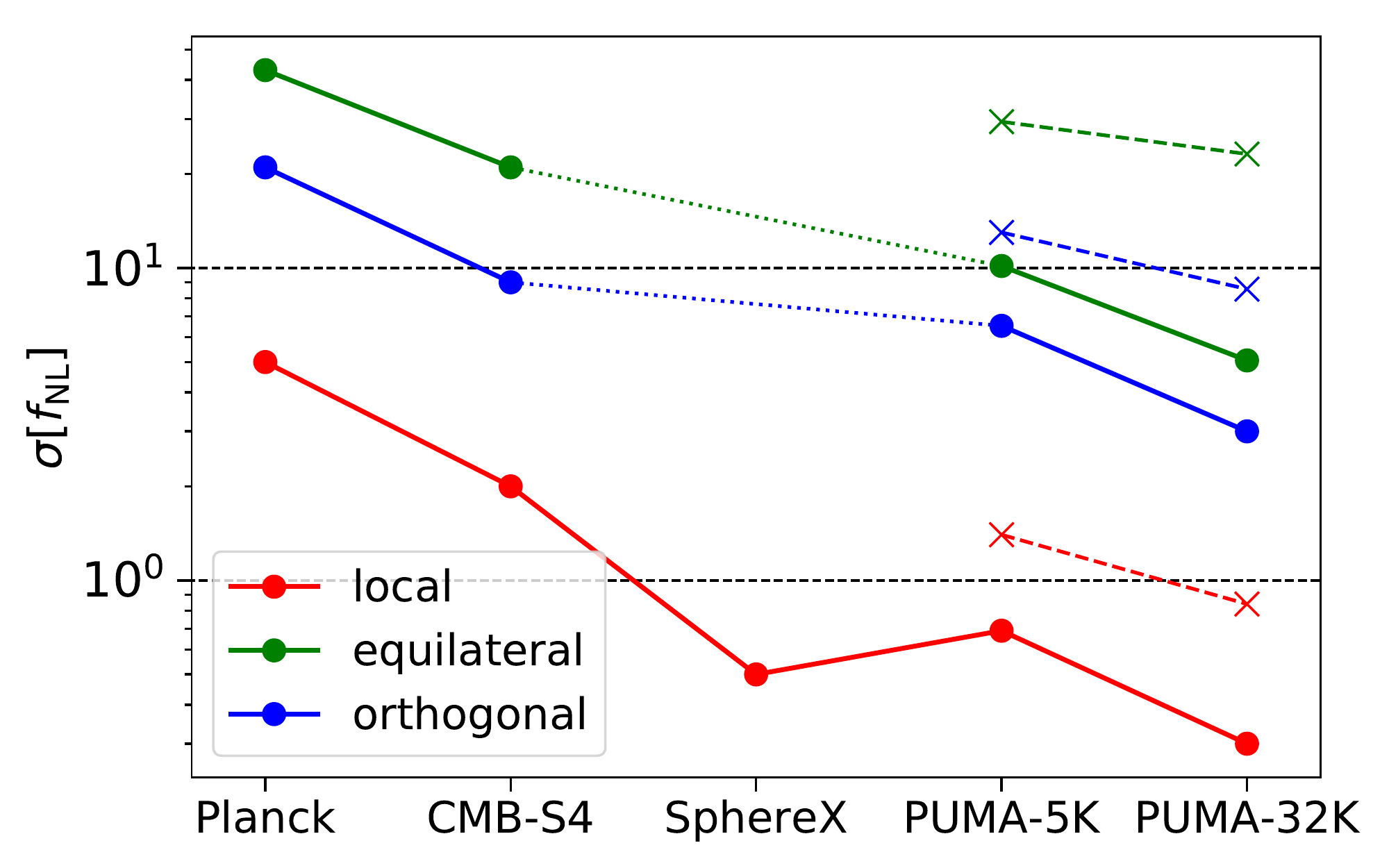} &
\begin{minipage}{0.6\linewidth}
  \small
  \vspace*{-5cm} 
  \begin{tabular}{|c|C{1.4cm}|C{1.4cm}|C{1.8cm}|C{1.8cm}|}
    \hline
    \multirow{2}{*}{$\fnl$} & \multicolumn{2}{c|}{CMB error}
    & \multicolumn{2}{c|}{PUMA error} \\
    &Planck (current) & CMB-S4 (forecast)& PUMA-5K & PUMA-32K \\
    \hline
    Squeezed (local) & 5.0 & 2.0  & 0.7 (1.4)  & 0.3 (0.8)    \\
    Equilateral &43 & 21 &  10 (30)  & 5.0 (23)\\
    Orthogonal & 21 & 9.0 & 6.5 (13)  & 3.0 (8.5)\\
    \hline
  \end{tabular}
\end{minipage}
  \end{tabular}
  \caption{$1\sigma$ constraints on various types of $\fnl$ parameters
    for PUMA as compared with the results from other experiments.  The
    table on the right shows the numerical values compared to current
    \cite{Stage2WhitePaper} and forecasted CMB errors.  Numbers in
    parentheses are for pessimistic foreground modelling.  We see that
    even the pessimistic foreground case results are competitive with
    other experiments and would be a significant step towards a
    characterization of the inflationary mechanism. The figure on the
    left shows the same information plotted graphically. Dashed lines
    are for the pessimistic foreground case. Dotted green and blue lines
    indicate that SphereX will not be sensitive to equilateral and orthogonal
    non-Gaussianity shapes.}
\label{fig:fnl}
\end{figure}

In addition to measuring parameters associated with how the Universe expands and evolves, PUMA would allow us to probe enormous cosmological volumes in order to constrain the mechanisms by which the primordial fluctuations were generated \textbf{(C)} -- currently posited to be a period of nearly exponential expansion in the very early Universe known as inflation.  The plot and table in Fig.~\ref{fig:fnl} show an example, where forecast constraints on the degree of non-Gaussianity in the primordial perturbations are compared.  Primordial non-Gaussianity is one of the very few observational handles we have on models of inflation, and reaching the critical sensitivity of $\sigma(f_{NL})\sim 1$ would have enormous implications for early Universe physics \cite{2019arXiv190304409M}.  We see from Fig.~\ref{fig:fnl} that while future CMB experiments struggle to reach the critical $\sigma(f_{NL})\sim 1$ threshold, PUMA has the statistical power to reach this level even under relatively pessimistic assumptions about foreground contamination.

\begin{figure}
  \centering
  \begin{tabular}{cc}
    \hspace*{-0.4cm}
    \includegraphics[width=0.52\linewidth]{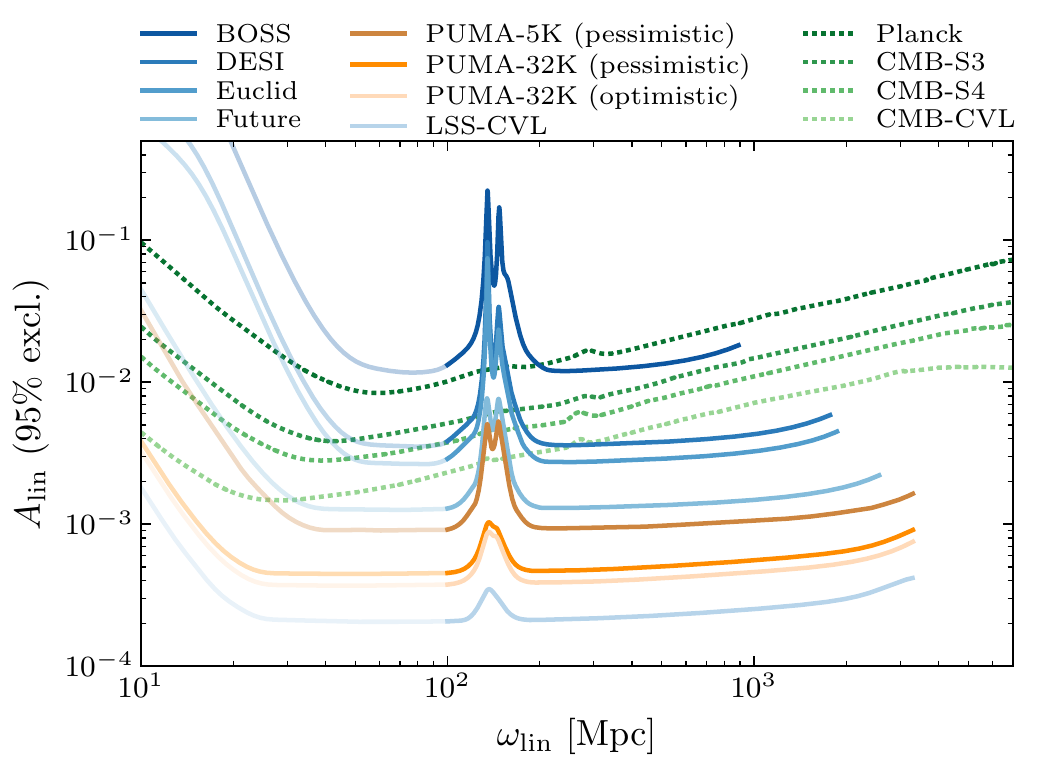}&
    \begin{minipage}{0.45\linewidth}
    \vspace*{-6cm}
    \small
    \begin{tabular}{c|cc}
        Frequency & PUMA-5K  & PUMA-32K \\
        \hline
        \SI{200}{Mpc}  & $0.8\ (0.9)\times\num{e-3}$  & $3.9\ (4.7)\times\num{e-4}$ \\
        \SI{2000}{Mpc} & $1.1\ (1.2)\times\num{e-3}$  & $5.3\ (6.5)\times\num{e-4}$ \\
      \end{tabular}
    \vspace*{0.1cm}

    Projected upper limits at 95\% C.L.\ on the amplitude of a linear feature with a given frequency. Values based on pessimistic foregrounds are provided in brackets.
    \end{minipage}\\
    \end{tabular}
    \caption{Left: Forecasted upper limits at 95\% c.l.\ on the amplitude of primordial features with linearly-spaced oscillations as a function of their frequency (see~\cite{Beutler:2019ojk} for details). Right: The numerical values of the sensitivity at two representative frequencies.}
\label{fig:features}
\end{figure}

PUMA can not only constrain cosmological higher-point correlation functions, but is also sensitive to primordial features in the two-point function~\textbf{(D)}. While models of inflation that invoke a single scalar field slowly rolling down a featureless potential produce smooth primordial spectra, there are numerous models with more complex potentials or multiple fields that can produce structure in these spectra~\cite{2019arXiv190309883S}. Figure~\ref{fig:features} illustrates the sensitivity of PUMA to constrain the amplitude of linearly-spaced oscillatory features in the primordial power spectrum as a function of their frequency (under the assumption the true amplitude is zero). The discovery of such features would provide much needed information on the inflationary epoch, and tight upper limits in their absence rules out various mechanisms for realizing inflationary models in theories of fundamental physics. PUMA could put upper limits on these oscillatory features a factor of 5 to 10 times better than current and near-term surveys, with PUMA-5K being more sensitive than a spectroscopic follow-up of LSST galaxies and PUMA-32K getting close to the cosmic variance limit.

\begin{figure}
  \centering
  \begin{tabular}{cc}
    \hspace*{-0.4cm}
    \includegraphics[width=0.6\linewidth]{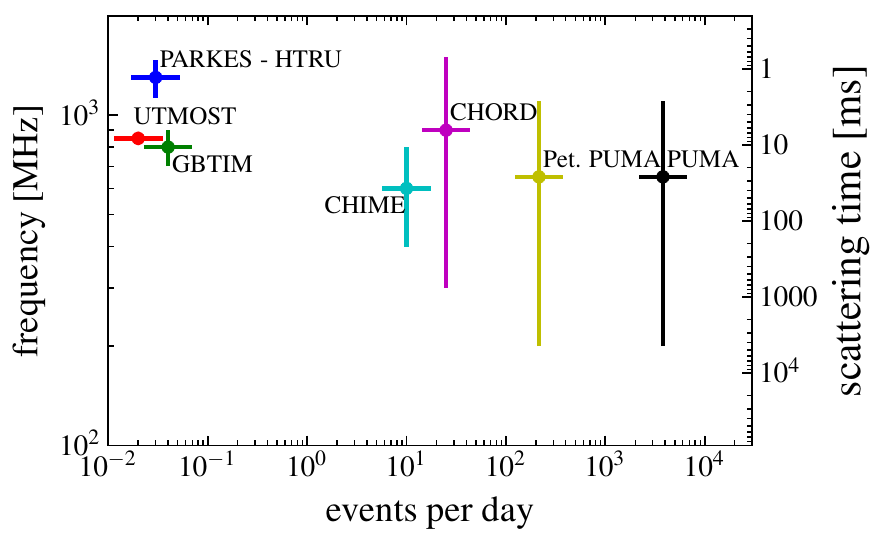}&
    \begin{minipage}{0.45\linewidth}
    \vspace*{-6cm}
    \small
    \begin{tabular}{c|cc}
        Frequency & PUMA-5K   & PUMA-32K  \\
        \hline
		  200-400\,MHz & 70 & 1200 \\
		  400-700\,MHz & 60 & 1000 \\
		  700-1100\,MHz & 80 & 1300 \\
      \end{tabular}
    \vspace*{0.1cm}

    Number of detected FRBs per day in three frequency bands.
    \end{minipage}\\
    \end{tabular}
    \caption{Left: The forecasted daily rate of FRBs and relevant frequency coverage for a selection of current and future experiments.   Right: The forecasted number of detected FBRs per day split into frequency bands.}
\label{fig:frbs}
\end{figure}

Fast radio bursts \textbf{(E)} offer a unique probe of the distant Universe if a
survey with large numbers of precisely localized sources is
available \cite{2019arXiv190306535R}. Faraday rotation provides a
precision probe of intergalactic magnetic fields and time-delay
microlensing allows for a cosmic census of compact objects (including
constraining black holes as dark matter). In addition, dispersion can
be used to locate the missing baryons \cite{2014ApJ...780L..33M,1901.02418}
and to measure the free electron power spectrum, breaking a degeneracy for
interpretations of kinetic Sunyaev-Zeldovich~(kSZ)~measurements.
The forecasts for the expected number of FRBs remain highly uncertain.
Nevertheless, under plausible assumptions about their propreties,
normalized by the CHIME rate of 5 per day, PUMA-5K should see around 215
FRBs per day while PUMA-32K's rate will be a full 3,500, or around two per minute.
  We illustrate and compare this with other current and planned experiments in Figure \ref{fig:frbs}.
  The methodology employed in this forecast, and primary references,
can be found in \cite{Stage2WhitePaper}.

The SKA1-LOW and SKA1-MID arrays will detect over 10,000
pulsars~\cite{2015aska.confE..40K}. It is clear that none of the
current telescope facilities, including SKA itself, would have enough
sky time to follow up the majority of these discoveries.

Due to the daily monitoring \textbf{(F)} of a significant subset of these pulsars
(depending on the pointing), PUMA will be complementary to SKA, even
in the petite array configuration. To see this, we note that PUMA has a significantly larger collecting area than the current SKA incarnations. In concrete numbers, the collecting area of PUMA is 0.14\,km$^2$ and 0.9\,km$^2$ for PUMA-5K and PUMA-32K, while the number of SKA-1 MID is 0.033\,km$^2$ and $\sim 0.1$\,km$^2$ for SKA-LOW at 300\,MHz \cite{1711.01910}. In addition, PUMA has larger bandwidth and roughly hour-long dwell times per transit. Therefore, as a first approximation we assume that if a pulsar is detectable with SKA, it should be within sensitivity reach of PUMA, even in the 5K configuration. 

We then estimate the total number of pulsars in the PUMA field of view as follows: we take the existing list of pulsars from the ATNF pulsar \cite{2016yCat....102034M} catalog containing 2704 pulsars and add a list of SKA1-LOW simulated pulsars containing 3036 pulsars for a total list of $5704$ pulsars \cite{ShiDaiPrivate}.  Note that SKA1-MID should find approximately 10,000 additional pulsars, many of which will be observable by PUMA, so our numbers are conservative.

\begin{figure}
    \hspace*{-1.5cm}
    \begin{tabular}{cc}
    \includegraphics[height=0.38\linewidth]{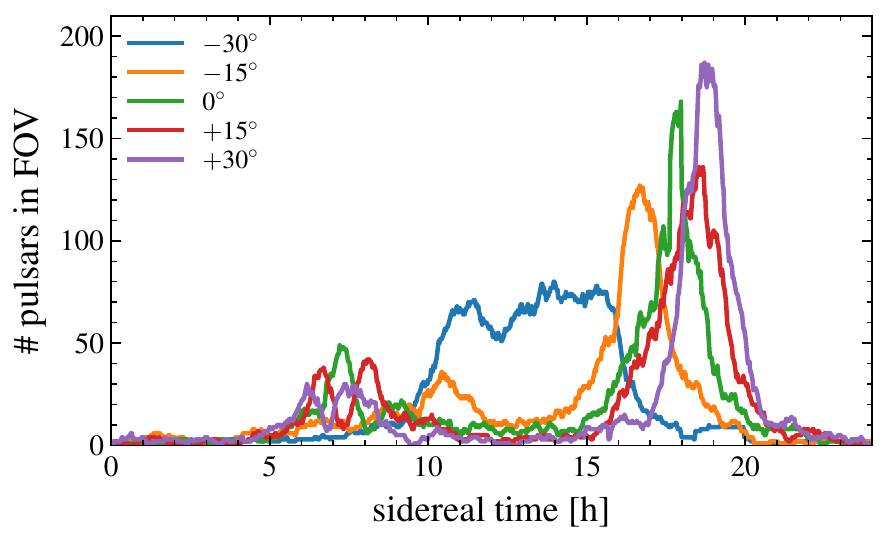} &
    \includegraphics[height=0.35\linewidth]{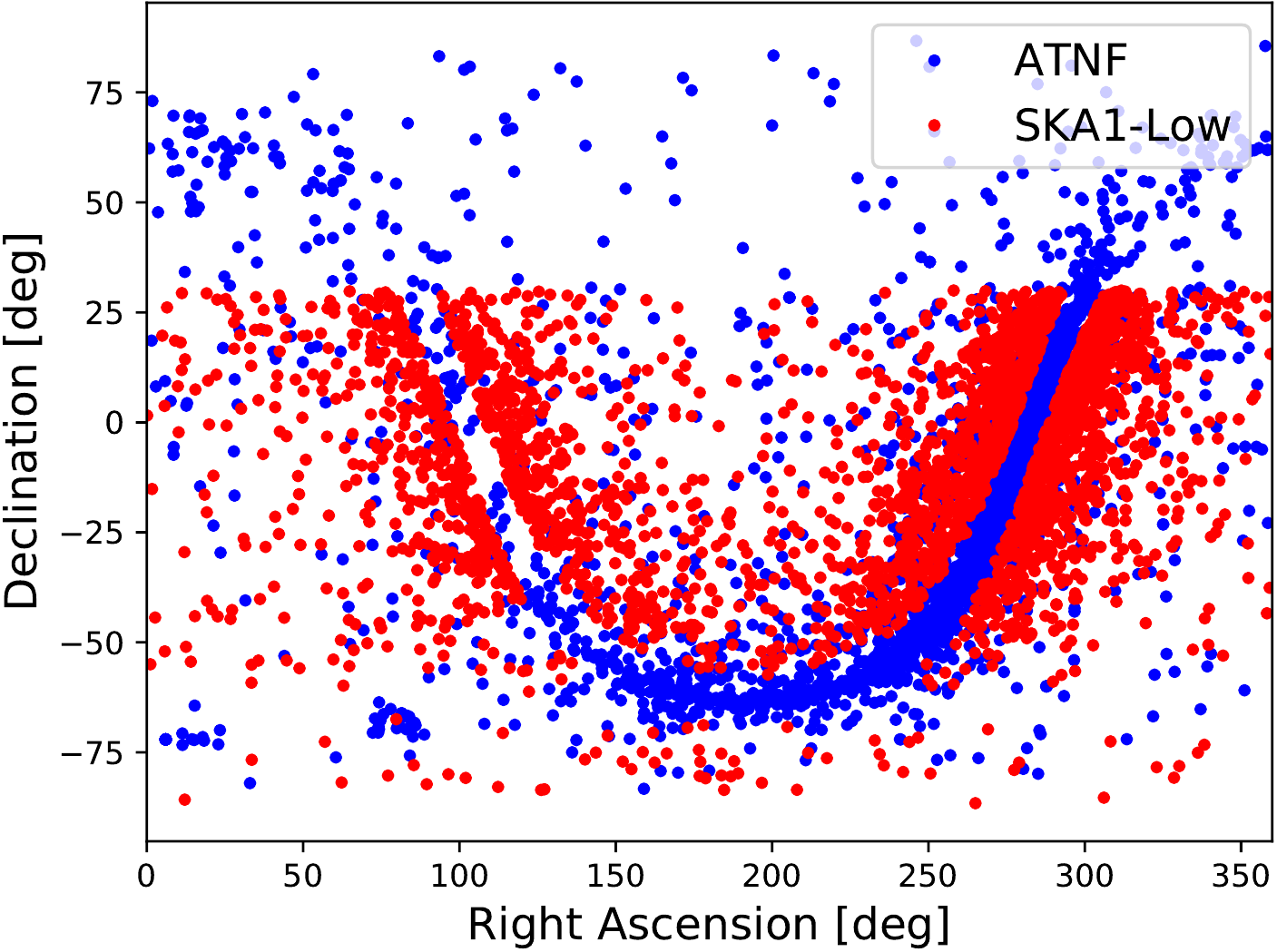}
    \end{tabular}
    \caption{Left panel shows the number of instantaneously observable pulsars in our catalog that is a combination of the real ATNF catalog and synthetic catalog of SKA1-Low pulsars. The right figure shows the distribution of these sources on the sky. SKA1-Mid is expected to find around additional 10,000 pulsars which were not used in these forecasts, but most of them should also be observable.}
    \label{fig:pulsarpr}
\end{figure}

We then assume PUMA to be observing from the latitude $-30^\circ$ (coinciding with the SKA observatory) and calculate the number of sources in the beam as a function of right ascension and offset from zenith pointing. This is plotted in Figure \ref{fig:pulsarpr}. 

Depending on the declination, the number of pulsars that are observable in a sidereal day is 397, 512, 663, 731 and 769 for observing at altitudes $-30^\circ$,$-15^\circ$,$0^\circ$,$+15^\circ$ and $+30^\circ$ from zenith for a total of 3072 pulsars or approximately half the total projected number of SKA pulsars. Of course, these numbers should be taken with appropriate uncertainties. For example, the size of the beam has been estimated at 500\,MHz and no account has been taken for beam suppression, etc. However, it is nevertheless correct to say that an instrument like PUMA is in principle capable of observing approximately 10\% of SKA pulsars daily with the total number surveyed over the nominal survey reaching up to 50\% of SKA pulsars.

In terms of time-of-arrival precision, our large collecting area, ultra-wide bandwidth, and
long contiguous observations (roughly hour transits across the 15-degree
beam, depending on observing frequency) will result in timing that is limited
principally by pulsar intrinsic red-noise timing jitter, which can be better
than 100\,ns in a single epoch
for certain millisecond pulsars. PUMA's wide band will allow a precise
measurement
of pulsar dispersion at each observation, which is crucial since time-variability in the
dispersion is a leading source of timing noise.

Another important consideration in pulsar timing is observing cadence, which is dictated by how often we
repoint the array to different observing altitudes. While our detailed survey
strategy has not yet been set, repointings will likely be in the range of months. As such, observing cadences for a given pulsar will be
around $\sim$1/hour/day for a month twice a
year. In any case, there will be significant months-long periods missing in the timing campaign.  The precise observing strategy is subject of further research.
Regardless of these details, our high sensitivity,
daily cadence during a given pointing, and a large number of monitored sources
will make PUMA the leading timing
telescope. It will be able to characterize
each of these new pulsar discoveries, and carry out a systematic study
of pulsar temporal variabilities, including nulling, glitches,
sub-pulse drifting, giant pulse emission, and potential signatures of
new fundamental physics.

	\begin{table}
		\scriptsize
		\begin{tabular}{l|l|l}
			\hline
			\textbf{Science Objective} & \textbf{Measurement Requirement} & \textbf{Instrument Requirements} \\
			\hline
			\textbf{A.} Characterize expansion history   & Cover $2<z<6$ and $k<0.4\hMpc$ & Bandwidth must include 200-475\,MHz\\
			in the pre-acceleration era      & with SNR per mode $\sim 1$ at $k\sim\SI{0.2}{\hPerMpc}$  & Maximum baseline $L_{\rm max}\gtrsim\SI{600}{m}$ \\
			\emph{Decadal Science Whitepaper: \cite{2019arXiv190312016S}}  &  & $ND>\SI{25}{km}$ at $L_{\rm max}=\SI{600}{m}$  \\
			\hline
			\textbf{B.} Characterize structure growth & Cover $2<z<6$ and $k\sim \SI{1.0}{\hPerMpc}$ & Bandwidth must include 200-475\,MHz\\
			in the pre-acceleration era      &  with SNR per mode $\sim 1$ at $k\sim \SI{0.6}{\hPerMpc}$ & Maximum baseline $L_{\rm max}\gtrsim\SI{1500}{m}$ \\
			\emph{Decadal Science Whitepaper: \cite{2019arXiv190312016S}}                            &  & $ND>\SI{200}{km}$ at $L_{\rm max}=\SI{1500}{m}$  \\
			\hline
			\textbf{C.} Constrain or detect primordial     &                             Measure $\gtrsim 10^9$ linear modes & Same as above plus:    \\
			non-Gaussianity &      with SNR per mode $\sim 1$  & bandwidth $200-1100$\,MHz ($z\sim 0.3-6$) \\
			\emph{Decadal Science Whitepaper: \cite{2019arXiv190304409M}}                            &                    &  assuming $f_{\rm sky}\sim 0.5$       \\
			\hline
			\textbf{D.} Constrain or detect features  &   Sufficient forecasted power spectrum sensitivity & Same as above \\
			in the primordial power spectrum                           &	& \\
			\emph{Decadal Science Whitepaper: \cite{2019arXiv190309883S}}        &	&   \\
			\hline
			\textbf{E.} Fast Radio Burst Tomography  & \quad -- 1 million FRBs  &
			Fluence sensitivity threshold $ \lesssim 2.5 f_{\rm sky}^{2/3} \si{Jy.ms}$\\
			          & \quad -- covering two frequency octaves &   Provide real-time FRB back-end\\ 
			\emph{Decadal Science Whitepapers: \cite{2019arXiv190306535R,2019arXiv190307370S,2019arXiv190309224K,Lynch}} &\quad  & Provide baseband buffer with triggered readout\\
			\hline
			\textbf{F.} Monitor pulsars  & Detect all pulsars in current Field of View    & $10\,\sigma$ point source sensitivity $>\SI{10}{\mu Jy}$/transit\\
			\emph{Decadal Science Whitepapers: \cite{2019arXiv190308653C,2019arXiv190308194F,2019arXiv190306526L,Bower,Lynch,2019arXiv190307644K}} &  brighter than \SI{10}{\mu Jy}  & Provide real-time pulsar back-end \\
			\hline 
		\end{tabular}
		\caption{\small Instrumental requirements implied by the science drivers (see also Table 1 of \cite{PUMAWhitePaper}). All derived instrument parameters assume certain fixed system properties such as amplifier temperature, sky background and various efficiency factors as outlined in~\cite{Stage2WhitePaper,PUMAWhitePaper}.  $N$ is the number of elements and $D$ is their linear dimension, $L_{\rm max}$ is the size of the longest baseline.}
		\label{tab:STM}
	\end{table}

\begin{figure}
\centering
\includegraphics[width=\linewidth, trim=0 10 0 0]{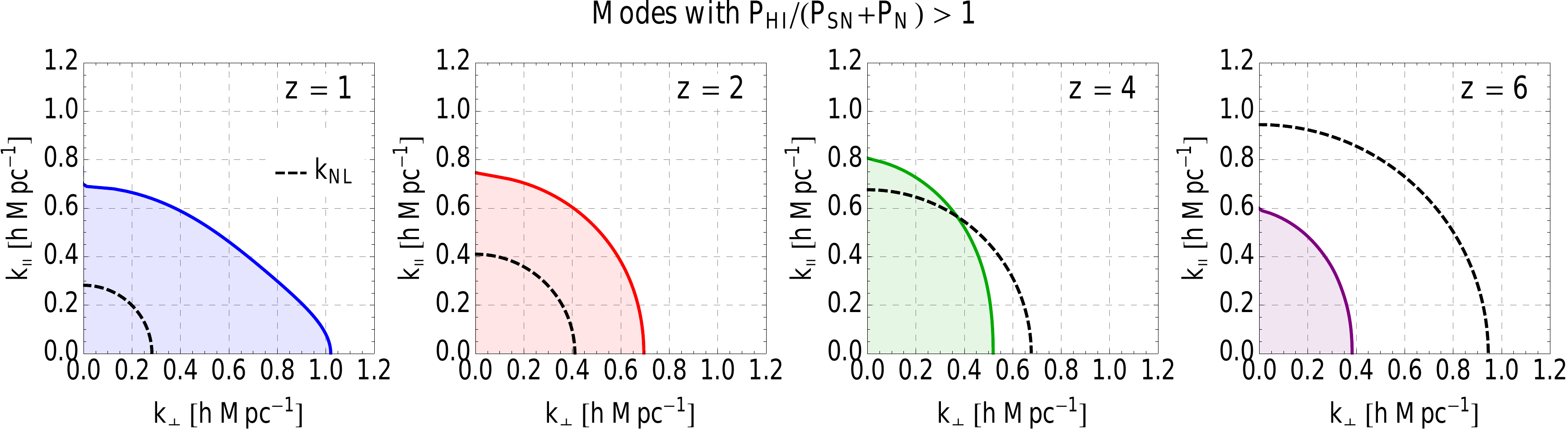}
  \caption{The shaded regions indicate which Fourier modes of the 21$\,$cm temperature field have ${\rm S/N}>1$ at a few representative redshifts for PUMA-32K, where the noise is a sum of thermal and shot noise. With such an instrument, ${\rm S/N}>1$ can be achieved for all linear modes at $z\lesssim 4$ and all modes with $k\lesssim 0.4h {\rm Mpc}^{-1}$ at $z\lesssim 6$. Foregrounds will of course reduce the number of accessible modes in practice, but will nevertheless leave a huge number of modes useful for cosmological and astrophysical studies.  Taken from \cite{Stage2WhitePaper}.}
\label{fig:snr}
\end{figure}

The instrumental requirements for these science cases are summarized in Table \ref{tab:STM} (adapted from \cite{PUMAWhitePaper}), with the SNR achieved shown in Fig.~\ref{fig:snr}.  For the IM science, the least demanding of our goals is (A), while the most demanding are (B), (C) and (D).  Goals (E) and (F) require additional back-end hardware.
Each of the science drivers translates into a natural required resolution (see table) which determines the longest baseline in the array.  The required sensitivity then sets the number of dishes times the linear dimension of each dish (see e.g.\ Appendix D of \cite{Stage2WhitePaper}).  The desire to probe the large-scale modes which reflect the primordial conditions of the Universe demands a close-packed array (since the smallest wavenumber that can be constrained, $k_{\rm min}$, is $\propto L_{\rm min}/\lambda$, the minimum baseline separation in units of the observed wavelength).

\subsection{Sensitivity of the Forecasts to Instrumental Parameters}

We did not study how all the forecasts respond in detail to changes away from the nominal configuration, but we did detailed trade studies for some science goals and all of our forecasts for both the full PUMA-32K configuration and for the smaller PUMA-5K configuration.
This leads to some intuition which we present in Table \ref{tab:psens}.

\newcommand{\imp}{$\circ$}
\newcommand{\vimp}{$\bullet$}

\begin{table}
  \centering
  \begin{tabular}{p{3cm}|cccp{6cm}}
    Science Goal & Longest B.  & Shortest B.  & Sensitivity &  Comments \\
    \hline
    A. BAO &    & \vimp & \imp &  PUMA-5K is sufficient for $low-z$ BAO. \\
    B. Growth & \vimp   & \imp & \vimp &  PUMA-5K is sufficient for low $z$ RSD, but high $z$ is more challenging. \\
    C. Non-Gaussianity & \imp   & \vimp & \vimp &  Losing short baselines hurts local non-Gaussianity. Foreground wedge is particularly problematic. \\
    D. Features & \vimp & \vimp & \imp & Ratio of longest-to-shortest baseline determines the search range. \\
    E. FRBS & \vimp &  & \vimp & Longest baseline will determine localization without outriggers. \\
    F. Pulsar search &  & & \imp & We find that for following up \emph{known} pulsars, PUMA-5K has sufficient point source sensitivity. \\
    \hline
    Reconstruction & \vimp & \vimp & \vimp & This method, described in \cite{Modi19b}, enables modes lost to foregrounds to be reconstructed, including those crucial for cross-correlation with surveys without radial sensitivity (LSST, CMB-S4). \\
    Cosmological parameters & & \vimp & \imp & Low $k$, linear modes which probe primordial physics are critical.
  \end{tabular}
  \caption{Sensitivity of science goals to main array parameters away from the PUMA-32K configuration for a few main array parameters: the longest baseline (determining resolution), the shortest baseline (determining the largest accessible scales) and sensitivity or equivalently baseline density. Full circle indicates high importance while  empty circle medium importance of keeping a given specification at PUMA-32K value. We caution that such tables always entail some oversimplification.}
  \label{tab:psens}
\end{table}

Each of these science goals responds differently to degradations in the instrument characteristics.  For example, our BAO forecasts are very robust to reduction in the maximum baseline as long as $L_{\rm max}>600\,$m.  This is because the BAO signal is localized to large ($100\,$Mpc) scales.  By contrast it is more sensitive to loss of small baselines, which reduce the sensitivity to modes near the fundamental frequency of the BAO, or to reduction in the sky coverage (which increases the sample variance of the measurement and is an important contributor to our error budget at low $k$).  The BAO signal can be robustly detected with even a fraction of the total collecting area, but the range of redshifts that can be probed is set directly by the instrument bandwidth.

Goals (B-D) and the measurements in Tables \ref{tab:params} and Figs.~\ref{fig:fsig8mg}, \ref{fig:fnl}, \ref{fig:features} are less forgiving.  Especially for the highest redshifts, the increased sensitivity of PUMA-32K is beneficial.  Again we show in each of the figures how the constraints on key cosmological parameters change when we consider PUMA with 5K dishes compared to 32K.  In this comparison the density of the array is held fixed, but its side length is changed thus reducing the highest achieved resolution and sensitivity. We note that non-Gaussianity is particularly sensitive to this change. For example, losing short baselines severely reduces the ability to measure non-Gaussianity of the local type, while leaving the orthogonal and equilateral types largely intact. On the other hand, inability to deal with foreground wedge would will hurt those two combinations beyond the loss in sensitivity, because the number of possible bispectrum triangles drops drastically in that case.

In all cases, the science reach of PUMA is intimately tied to its ability to remove the large foreground signal. This a technical rather than fundamental problem, but instrumental design choices are vital to support it. Foreground mitigation requires exquisite instrument calibration and characterization.  The forecasts presented here make optimistic or pessimistic assumptions about foreground subtraction or mitigation so as to show the sensitivity to this issue, but even our ``pessimistic'' foreground case has yet to be demonstrated on real data.  Our pessimistic case assumes a three times primary beam, rather than horizon, foreground wedge. We find that if only modes outsize the horizon foreground wedge are used, PUMA is not competitive with other surveys for several science goals, therefore it is crucial that the instrument enables at least some level of foreground cleaning (rather than avoidance only).  It is believed that foreground control is easier with bigger dishes, because it is easier to suppress sidelobes. On the other hand, bigger dishes necessarily imply an increase in the shortest baselines. The needs of science for short baselines will have to be carefully stacked against the needs to have solid systematic control. These issues will be discussed in further detail in the following sections.

\subsection{Comparison with other radio telescopes in terms of Science}

Both configurations of PUMA are highly complementary to other planned and proposed large radio telescopes. The Next Generation Very Large Array (ngVLA~\cite{2018ASPC..517....3M}) does not overlap with PUMA’s frequency band and has entirely different scientific goals.

The Deep Synoptic Array (DSA-2000~\cite{hallinan2019dsa}) is in many aspects a similar instrument: several thousand dishes that harness advances in digital signal processing to reduce data in real-time. However, it is optimized with different science goals in mind, on a shorter timescale and for a different community.  It operates at $0.7-2.0\,$GHz, is fully tracking and it is not in a fully compact configuration,  which makes it less suitable for intensity mapping. In particular, even though it in principle reaches $z\sim 1$ in terms of the 21\,cm line, intensity mapping is not a primary DSA-2000 objective. PUMA will have a higher cost and spend significantly more effort in making the system extremely stable, with elements that are repeatable, characterizable, and calibratable to high precision, which will enable the intensity mapping application. For transient science applications, the two instruments will be complementary, both in frequency and sky coverage.

While SKA1 will undertake cosmological measurements at $z<6$~\cite{Bacon:2018dui}, its design must also satisfy the requirements of a much broader portfolio and science cases. PUMA therefore offers a number of distinct advantages in constraining cosmology. The radio continuum surveys planned for SKA1-MID are expected to identify a total of $\sim 1.6\times 10^8$ galaxies, which can at best be sorted into 5-10 broad ($\Delta z \sim 0.3$ - 0.5) redshift bins~\cite{Bacon:2018dui}, while PUMA’s intensity mapping dataset will be equivalent to a {\em spectroscopic} survey of $\sim 5\times 10^8$ galaxies for PUMA-5K or $\sim 2.5 \times 10^9$ for PUMA-32K. (SKA1-MID will also carry out an HI redshift survey at lower redshifts,  $z\lesssim 0.4$, mostly inaccessible to PUMA.) SKA1-Low’s HI intensity mapping survey at $3<z<6$ will be limited to roughly 100$\,{\rm deg}^2$, while SKA1-MID’s IM survey over $0.35<z<3$ will only use dish auto-correlations due to the lack of the short baselines most useful for large-scale cosmological measurements such as BAO~\cite{Bull:2014rha}. As a result, PUMA’s cosmological constraints will be much improved compared to SKA1: for example, even PUMA-5K will constrain the time evolution of $f\sigma_8$ more precisely (compare Fig.~\ref{fig:fsig8mg} of this document with Fig. 11 of~\cite{Bacon:2018dui}).

SKA2 will potentially have a large cosmological reach; however, as the specifications for SKA2 have yet to be fully determined, a direct comparison is difficult. In any case, it is likely safe to assume that SKA2 will be a general purpose instrument in the same manner as SKA1 will be. Moreover, one expects all the aforementioned instruments to learn from each other; for instance, high dynamic range imaging techniques developed for one instrument will almost certainly be applicable to the others if appropriately adapted.

Finally, the epoch of reionization arrays, such as HERA, aim at fundamentally different science.  By attempting to measure the signal at redshifts beyond six, they are interested in astrophysics of reionization, rather than fundamental cosmological questions. Nevertheless, in terms of controlling systematics, their challenges largely overlap with post reionization experiments such as PUMA and its precursors. We expect significant  cross-fertization of ideas and solutions to calibration problems.

\section{Enabling Technologies}

There are two crucial pieces of technology that exist today, but did not exist two decades ago, and which enable development of experiments like PUMA:

\begin{itemize}
\item \textbf{Commercialization of high-performance radio-frequency
    electronics} driven by the telecommunications industry continues to decrease the price points and availability of not only ADCs with giga-sample/s sampling rates and large dynamic range, but also the entire ecosystems of digital processing platforms that enable subsequent processing at low cost and low power consumption;
\item \textbf{Advances in hardware for high-throughput computing}, mostly through GPUs that were initially advanced by the gaming industry and, more recently, by the explosive growth in artificial intelligence.  The latter, in particular, has also led to the development of software platforms upon which we can build scalable, high performance code.
\end{itemize}

These two crucial technologies enable the telescopes to be built with
electronics taking the place of optics and analog signal chain,
bringing about massive cost-efficiencies and improvements in
systematics control.  The key telecommunications and digital computing
technologies on which the project depends are readily available in the
US.

These arguments are not new and have in fact led to experiments like CHIME and HERA. What these experiments have taught us, however, is that while making telescope capable of astronomical observations is relatively straightforward, reaching the required dynamic range and stability is considerably more difficult than anticipated even a decade ago \cite{aguirre2019roadmap}. This cautionary tale is central to our approach.

Our plan is thus to develop and demonstrate all the necessary technologies and refine the design requirements before we commit to metal. This effort will be four-pronged: \textbf{R\&D prong \#1} is the laboratory validation and demonstration of the individual pieces of technology, such as clock distribution, which we discuss in Section \ref{sec:hardware}. \textbf{R\&D prong \#2} will be extensive computer simulations of the entire system, including electromechanical properties of small clusters of individual receiving elements, which we discuss in Section \ref{sec:computing}. \textbf{R\&D prong \#3} will be detailed study into algorithms and performance of the real-time data reduction and calibration, discussed in Section \ref{sec:realtime}. Finally, \textbf{R\&D prong \#4} will be small-scale path-finder arrays discussed in Section \ref{sec:pathfinders}.

This approach is driven by several considerations. First, the systematic error budget will necessarily have to be distributed between various subsystems and preliminary work will inform various trade-offs. For example, we know that we will likely need sub-picosecond  clock synchronization across the array, but whether the requirement is 100\,fs (femtoseconds) or 1\,fs remains to be determined through careful computer study. Small pathfinder arrays can teach us about the overall system primary beam, phase and gain stability, but are of little use for foreground wedge cleaning (insufficient information) or real-time calibration (insufficient signal-to-noise). However, measurements from path-finder arrays can be used to inform computer simulations which can in turn inform the feasibility of real-time calibration and foreground control. Finally, due to programmatic reasons it is unlikely that any project funding would be available before 2022 and so all R\&D will need to occur under the umbrella of general non-project-specific research and detector development.

\section{R\&D prong \#1: Technology Development in the Laboratory}
\label{sec:hardware}

\subsection{Optical and Mechanical Antenna Design Optimized for Intensity Mapping (2021-2024)}
\label{sec:optic-mech-antenna}

We will need to undertake a study optimize the optical and mechanical designs for the individual interferometric elements to maximize sensitivity for the PUMA science goals. There are two aspects that are essential in this optimization: the wide bandwidth and the need for exquisite primary beam stability.

Feed antennae capable of our required bandwidth ratio of over 5:1 have been demonstrated in a dual polarization setup at high coupling efficiency ($>90\%$) across the frequency band\cite{Vanderlinde}. For example, DSA-2000 proposes a bandwidth of 700-2000MHz (1:3), while  the CHORD experiment uses 300-1500MHz (1:5). One of the potential issues is that current designs are rather large and present significant obstruction of the reflector if positioned at the focus. This may decrease the collecting area efficiency, and is likely to scatter radiation and increase spurious coupling between nearby dishes. Designs of both feed antennae in isolation, as coupled to the dish, and of various baffling schemes are part of this research. 

The other significant requirement in dish design is to ensure uniformity, stability and spectral smoothness of the final electromagnetic response. Designs will likely involve low-$f$-number, fast telescopes, which allow the primary focus to ``hide'' below the edge of the parabolic surface, thus minimizing the coupling between elements. The optics around the primary focus need to minimize reflections while also allowing for ultra-wideband operation. The design should allow off-zenith operations while maintaining the closest possible baselines and minimizing the optical shadowing of adjacent elements. The response of individual elements should change slowly with temperature and have predictable gravitational deformation, all while maintaining an excellent cost-efficiency as required for this experiment.

Finally, the differential response to linear polarization components needs to be simulated and well-controlled. The 21\,cm signal is unpolarized while many foregrounds are polarized.  Unlike the simplest models for foreground contamination, polarized foregrounds have a complicated structure as a function of frequency and could contaminate the real signal.  This structure arises because the polarized foregrounds are rotated by the interstellar medium on their way to the instrument by Faraday rotation, a frequency-dependent process that depends on the electron density and magnetic fields integrated along the line of site (Faraday depth).  Further, as noted in \cite{Shaw:2014vy}, this effect is a function of position in the beam.  Our dishes will need to have well understood polarization response (even when the dish is tilted), the requirements on which will be set iteratively by pipeline simulations.

It is clear that a careful design will be a major undertaking. We plan to achieve these with a combination of computer simulations (see Section \ref{sec:EMSim}) and construction and characterization of actual prototypes (see Section \ref{sec:pathfinders}).

\subsection{Dish Construction Techniques (2021-2028)}
\label{sec:dishConstruction}

Uniformity and stability of individual antennae is of crucial importance for calibration and efficient data processing and compression. For an array the size of PUMA, a study must be undertaken to assess how to produce the required number of uniform dishes in a cost-effective way, and how to verify uniformity. It is almost certain that the facility to produce and characterize main dish surfaces will need to be co-located with the experiment.

Significant progress in using carbon fiber composite dishes for repeatable dish shapes has been made for the SKA and cheaper fiberglass alternatives are being prototyped for the future 21\,cm arrays HIRAX and CHORD. Additional development will be required to fabricate dishes at PUMA scale within cost using any of these materials. A study to determine the trade-off between achieved surface errors and per-unit cost will be part of the development stage. At a minimum, experience and testing from current and upcoming experiments will inform this study, and may also drive us towards developing alternatives (e.g.\ fiberglass reinforcement, hard foam, etc) to meet specifications.

\subsection{Integrated Analog and Digital Front-end (2020-2024)}
\label{sec:integratedFrontEnd}

Although digital RF electronics were already well-developed for radar, military, and satellite communications by the early 1990s, an explosion in capability has been brought about by the advent of mass-market wireless applications: cellular telephony, digital broadcast TV, WiFi, Bluetooth, etc. This has led to the widespread availability of commodity components and systems, which are now being exploited for radio astronomy. In addition, market demand for higher-performance computing (FPGAs and GPUs for AI, machine learning, gaming, etc.) is contributing to the dramatic erosion of prices for state-of-the-art microelectronics. One example is the Xilinx RF-System-on-Chip (RFSoC) product portfolio \cite{farley2018all}, integrating giga-sample/sec digitizers and powerful $14\,$nm CMOS FPGA, DSP, CPU, and memory resources targeted at the wireless communications industry. Second, third, and fourth-generation parts with higher performance and lower power consumption are already in development. There are concurrent developments in analog low noise amplifiers, filters, and gain blocks well-suited to the frequency bands of interest for PUMA.

These developments have motivated early-stage research into compact receiver front ends for radio astronomy arrays \cite{Hampson19,CasperZcu111}. These technology developments make receiver electronics with integrated digitizers, digital filters, frequency translation, equalization, channelization, RFI mitigation, fast transient triggers, and high speed optical I/O an attractive option for PUMA. By digitizing at or near the focus, analog component performance requirements are relaxed, and long-distance data transport is more robust and easily interfaced to commodity switches and servers for correlation operations.

\begin{figure}
  \centering
    \includegraphics[width=\linewidth]{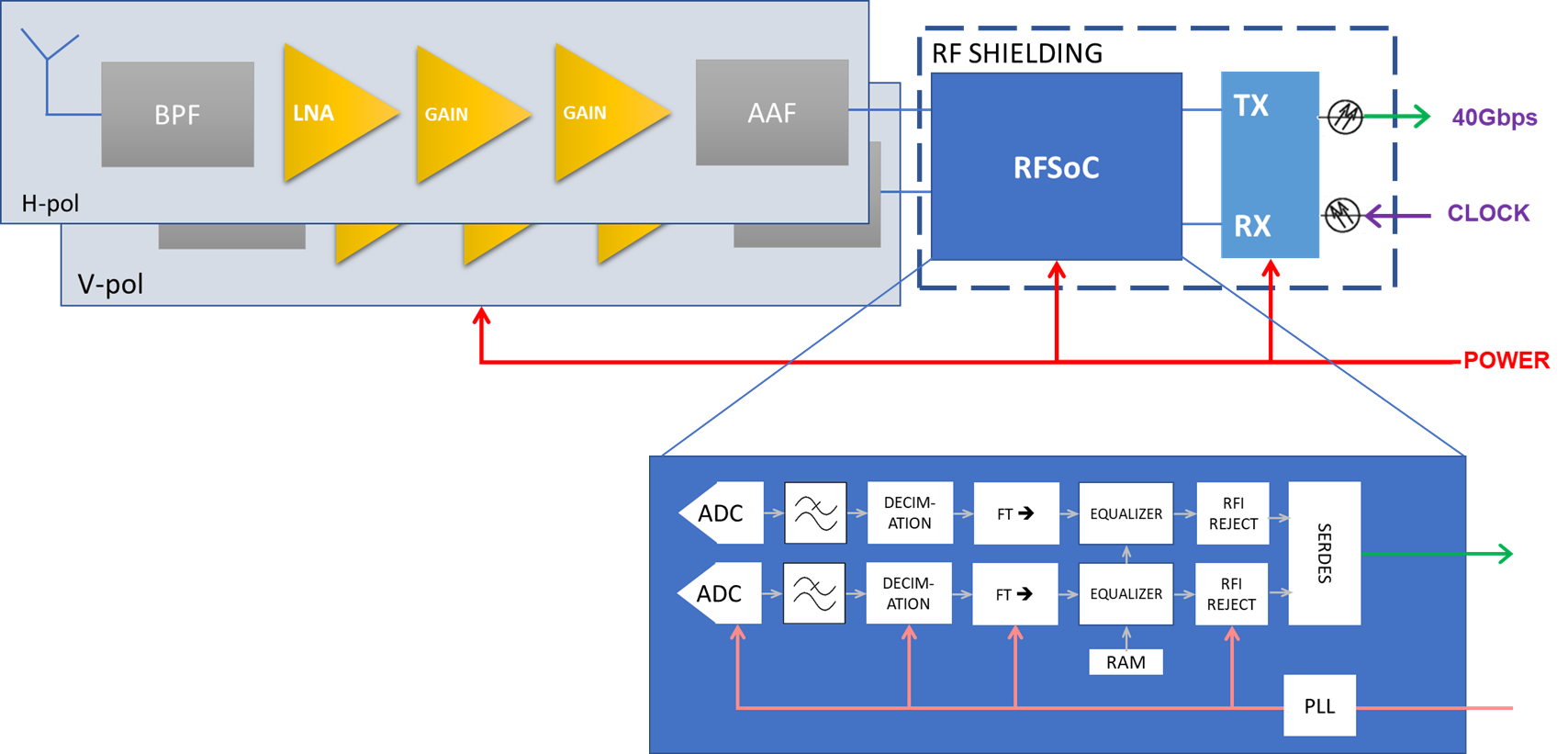}
\caption{Simplified block diagram of receiver electronics.}
\label{fig:frontend}
\end{figure}

A simplified block diagram of a compact receiver with digitization is shown in Fig.~\ref{fig:frontend}.  Early investigations will be targeted at measuring the effectiveness of RF shielding to prevent digital noise coupling into the signal path. If sufficient isolation cannot be obtained, the digitizer/channelizer boards can be deployed in the well-shielded timing distribution huts envisioned to serve clusters of 6-8 antennae. With multi-channel RFSoC boards in these locations, analog signal transport would only span $\sim10\,$m and some inter-antenna operations could be performed; e.g.\ crosstalk correction, partial correlation, etc.

\subsection{Clock Distribution (2021-2023)}

Clock jitter between correlation components can decrease the correlation efficiency and thus impact signal stability. In our current design, the signal is amplified and digitized at the reflector focus in an integrated receiver/feed antenna. This approach simplifies the signal handling down-stream, but does not by itself solve the phase jitter stability problem, unless the clocks driving the ADC are synchronized to sub-ps precision, which is several orders of magnitude below the sampling interval. Conversely, if such synchronization is achieved, the entire system gains orders of magnitude in stability -- it does not matter if the data packets arrive somewhat later due to thermal expansion in the cables as long as they are correctly time-stamped. Ensuring proper synchronization is critical to PUMA's success, although additional methods of achieving phase stability may be explored.

Luckily, synchronized timing is critical to many experiments, and photonic phase synchronization systems (for example baselined for SKA-MID~\cite{1805.11455}) achieve $\sim 50\,$fs stability over minute timescales with direct application in the radio telescope domain, while more sophisticated versions claim $10^{-17}$ stability over hour timescales \cite{Wang:19}. The main R\&D effort will be to adapt and validate these techniques for our particular case while minimizing costs to stay in budget. In the current cost forecasting, we assumed the stabilized signal arrives at a base-station where it would be distributed over 6 dishes, but with sufficient development it should be possible to integrate this technology into the per-dish digital front-end described above.

\subsection{Primary Beam Calibration (2019-2023)}

Sobering experience from IM surveys over the past decade \cite{aguirre2019roadmap} points out the importance of beam calibration. Due to the variety of ways that foregrounds couple into the instrument response, it has become clear that better designs and improved characterization and modeling will be necessary for IM experiments to realize their promise. Coordinated investments in instrument, analysis, and software technologies across the IM community, including those proposed here, will have a large payoff in the next decade.

Preliminary simulations indicate we must know the beam to a precision of $\sim0.1\%$ \cite{Shaw:2014vy} across the entire band to enable adequate foreground removal. Precautions must be taken to avoid internal resonances or reflections that introduce non-smooth spectral structure into the instrument response. In addition, uniformity must be characterized in order to take advantage of redundant calibration schemes \cite{2010MNRAS.408.1029L} and co-adding of similar baselines for FFT-based correlation algorithms. If dish uniformity is confirmed to be within specification, then the beams of only a few dishes must be measured well. These beams can be compared to electro-magnetic simulations, and deviations in the beam can be assessed from the data itself (and any outliers confirmed with careful beam measurements), and should be parametrized easily with a few additional parameters. To quantify uniformity and ensure foreground removal, mapping of individual beams at sub-percent precision will be an important aspect of the calibration of the entire system.

Current-generation experiments are developing the calibration strategy for transit interferometers. Their approach uses a variety of measurements: the few bright celestial sources that can be reliably measured without confusion noise from other sources; measurements of the sun (with solar variability correction extrapolated from solar telescopes); pulsars (where their flashing allows removal of all confusion noise); and satellites (which are narrow band, and usually variable during a beam crossing)\cite{1505.07114}. It is expected that deeper maps which cover PUMA's wide frequency range and include extragalactic point sources, diffuse and polarized galactic emission will become available in the next several years, perhaps from dedicated mapping arrays. These maps will help anchor calibration methods tied to sky models.

\begin{centering}
\begin{figure*}[ht]
{\includegraphics[width=0.9\textwidth]{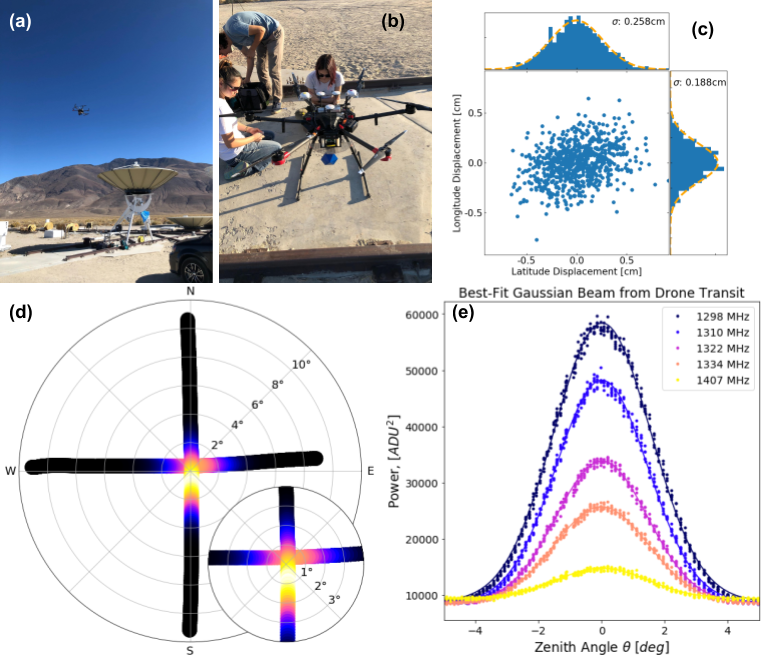}}
{\caption{\label{fig:drone_results} {\small \textbf{\textit{(a)}} A photo of the drone flying over 4.6m dishes at OVRO \textbf{\textit{(b)}} The drone outfitted with a biconical antenna (blue) and radio source \textbf{\textit{(c)}} At OVRO we assessed the drone's location accuracy while stationary: the results showed 19mm longitude and 26mm latitude accuracy \textbf{\textit{(d)}} drone flight trajectory, color scale indicates brightness of signal as measured by the OVRO dish \textbf{\textit{(e)}} beam slice in one dimension with Gaussian fit. }}}
\end{figure*}
\end{centering}

These experiments have also begun developing calibration sources deployed on autonomous aerial vehicles (see Figs.~\ref{fig:drone_results}, \ref{fig:bmx}). Application of quadcopter drones to beam mapping has been attempted by several groups (e.g.\ \cite{Chang_drones,Virone_drones,Jacobs_drones,Pupillo_drones}), with some preliminary success. Although R\&D is still required for improved sources, flight times, RFI mitigation, and location accuracy, it is a promising method for full 2D beam coverage and assessment of cross-talk, given the unique capability of flying drones over stationary dishes.

Current problems identified from a variety of groups include overcoming the dynamic range between the main beam and the lower sidelobes (sources bright enough to measure the sidelobes saturate the main lobe, reducing the ability to `stitch the measurements together' for accurate maps). One solution is to fly multiple times with a variety of attenuation levels, but this requires far greater flight times while groups are already struggling to produce simple slices with current flight times. In addition, contributions from the changing sky brightness will impact measurements at the 0.1\% level, for which flashing the source is a good approach but one that also requires additional flight and integration time. A variety of approaches are being considered, including digital calibration sources which may also be used for gain stabilization if digitization at the focus is successful (Yale and WVU) and fixed wing drones that are considerably faster and with longer flight duration (BNL). This latter approach allows scanning patterns that can cross-link, allowing the $1/f$ gain noise of the amplifiers to be backed out. The downside is that sufficiently fast and accurate positioning is difficult and that flight is less stable in windy conditions.

These techniques will continue to be developed over the next few years. Based on lessons learned from current-generation experiments and the understanding we hope to gain from PUMA R\&D, it is reasonable to expect that the technical hurdles can be overcome by the start of PUMA operations.

\section{R\&D prong \#2: Computing, Software and Pipelines}
\label{sec:computing}

Extensive computer simulations of all subsystems are used extensively in the particle physics community and are becoming more common in astrophysics (e.g.\ Planck, LSST, SKA).  Such techniques are particularly appropriate for the design and validation of the intensity mapping experiments, due to the inherently difficult data analysis that is at the same time under good theoretical control (i.e.\ we have a good understanding of Maxwell's equations and the radiation mechanism producing the source signal).

There are three main aspects of computing that naturally factorize. First there are electromagnetic simulations of the response of a set of closely-packed interferometric elements, which require specialized software packages and effort. Second is modelling of the timestream data, which is then analyzed with the third: algorithmically complete pipelines. We discuss these in the following subsections.

\subsection{Electromagnetic modelling of the array performance}
\label{sec:EMSim}

In the traditional literature on radio interferometers, individual
interferometric antennae are assumed to be independently sampling the
incoming radiation. However, since the cm-wave closely-packed
interferometers for CMB observations were built in 2000s
\cite{2001ApJ...549L...1P,2002ApJ...568...28L,2003MNRAS.341.1057W},
spurious correlations between nearby elements have been observed that
were above the level expected from amplifier noise coupling and were
never completely explained
\cite{2003MNRAS.341.1057W,2001ApJ...549L...1P}. More recently, early
results shows that current-generation instruments like CHIME and HERA
may suffer from similar ailments, although further work is required to
definitively ascertain their origin
\cite{ACLPrivate,1908.02383,1909.11732,1909.11733}.  Since such
effects can destroy our ability to perform foreground cleaning, it is
crucial that they are better understood. Fortunately there are now
multiple paths forward, whether they involve the large amounts of data
from current-generation instruments that are now available for study,
or via the PUMA-centric path outlined below.

Our plan is to understand such effects with extensive electromagnetic simulations of the entire system. This is a difficult computational problem with two main commercial providers of EM solvers: ANSYS HFSS and Dassault Syst\`emes' CST.  We plan to develop extensive simulations of our entire system including:

\begin{itemize}
\item Simulations of the electromagnetic properties of a single isolated dish. This should include the full mechanical and material properties of the system, including ortho-mode transducer (OMT) and support structures.

\item Simulations of the same under realistic manufacturing tolerances and geometric changes caused by environmental variations (temperature, humidity).

\item Simulations of a pair of receivers as a function of their relative positions, including cross-couplings, correlated ground pick-up and shadowing at off-zenith pointings.

\item Simulations of several receivers in our typical 50\% filled hexagonal closely-packed configuration to simulate the ``continuum'' limit of many dishes.
\end{itemize}

Where possible, we will validate these simulations against measurements made on the engineering pathfinder arrays (Section \ref{sec:pathfinders}).

It is not clear if the later stages are computationally feasible with the available software packages. In that case we will develop appropriate approximation methods.

The main purpose of this exercise is two-fold. The first is to gain insight into which properties of the receiver matter most and how the imperfections degrade the performance. This will inform various trade-off studies and help set-up needed tolerance requirements. The second is to provide a simplified ``effective theory'' of the dish array to be used in the signal time-stream modeling.

\subsection{Timestream simulation}

We anticipate that the analysis of PUMA data will require taking into account numerous decisions made in the process of data collection and processing.  The most complete means of testing the impact of such decisions, and eventually accounting for them, relies on an end-to-end simulation of the data-gathering through timestream simulation.  This is also the best means of conveying our knowledge of the data products as they relate to the input sky signal.  We envision building synthetic data sets for development and deployment, validation and verification and uncertainty quantification.

Starting from a model of the signal and the major noise sources, a timestream is produced by simulating the observational strategy and modeling the response of the instrument to the various time-dependent inputs.  Similar approaches are already commonplace in CMB experiments \cite{2018arXiv180706207P,2019arXiv190704473A} and are being developed for some epoch-of-reionization experiments and existing hydrogen intensity mapping experiments. This timestream is then processed by a (possibly simplified but) functionally complete analysis pipeline (next section) to produce synthetic data products which can be used as input to science analyses.

In our favor, while the challenges facing $21\,$cm surveys up to $z\simeq 6$ are significant, they are reasonably well understood.  The HI signal itself can be simulated quite efficiently using modern N-body techniques, even over the vast volumes that will be probed by PUMA (see e.g.~\cite{Modi19b} for one of several techniques that have been realized recently), and there are ongoing community efforts to calibrate these methods using hydrodynamic simulations and observations.  To this `signal' map are added galactic and extragalactic foregrounds, and a model of ionospheric effects, then the input is convolved with the complex beam for each antenna.  For a transit instrument there are significant performance gains that can be achieved in this step using existing tools (e.g.\ \texttt{driftscan}\footnote{\url{https://github.com/radiocosmology/driftscan}} or TOAST\footnote{\url{https://toast-cmb.readthedocs.io/en/latest/intro.html}}).  Time-dependent beam convolution for temporally varying sources (e.g.~solar, jovian and lunar emission) and the effects of RFI can be added straightforwardly.  Gain variations can be directly applied to the timestreams which are then calibrated and compressed.  We anticipate that we can break up many of these analysis steps, for example generating `effective' calibration uncertainties that can be applied directly to compressed timestreams.  The compressed data are then fed into the data analysis pipeline and propagated through to the final science results to assess analysis systematics, instrument design, real-time calibration and data processing.

The minimum inputs required for this process are sky maps (including the $21\,$cm signal, galactic foregrounds and extragalactic point sources), instrument characteristics and models of the observing conditions including RFI and the ionosphere.
Developing such simulations is a time consuming exercise that will need to be started early, however we also benefit from the fact that our understanding of the low-frequency sky is improving steadily from ongoing radio, CMB, \stageone\ $21\,$cm and epoch of reionization experiments and those communities are developing computationally efficient and scalable tools\footnote{A small sample of relevant software packages includes:\newline
PySM: \url{https://pysm-public.readthedocs.io/en/latest/index.html}\newline
CRIME: \url{http://intensitymapping.physics.ox.ac.uk/CRIME.html}\newline
CoLoRe: \url{https://github.com/damonge/CoLoRe}\newline
GSM: \url{https://github.com/jeffzhen/gsm2016}\newline
CORA: \url{https://github.com/radiocosmology/cora}
}.

\subsection{Pipelines}

Computing requirements for PUMA come from both the correlation burden and the data reduction, transfer, storage, analysis and synthetic data production.  Releasing science deliverables to the community from PUMA depends crucially on developing and deploying an analysis pipeline that can ingest vast quantities of data and transform it into well characterized frequency `maps' and power spectra.  This pipeline must pass an extensive set of validation and verification steps.
In order to facilitate trade-off studies, and to validate analysis codes and theoretical models, we need to begin building functionally complete pipelines for the PUMA instrument.

While an understanding of the final data products will require passing through all stages of the pipeline in an iterative fashion, refining the analysis and design, conceptually the pipeline can be divided into three broad areas:
\begin{itemize}
    \item Flagging, calibration and data processing including cleaning the data of RFI by flagging times and frequency channels that have been contaminated.  This is a well-understood problem within radio astronomy and the first generation experiments are already making good progress on this issue.
    \item Foreground removal, while preserving signal in as much of the Fourier plane as possible.  Even though foreground cleaning is a common problem in many areas of astrophysics and cosmology, the required dynamic range for $21\,$cm experiments is particularly challenging. Numerous methods have been discussed in the literature that allow for blind statistical separation of signals, including    principal component analysis (PCA), independent component analysis (ICA), generalized morphological component analysis (GMCA) and others \cite{1209.4769,2012MNRAS.423.2518C,2016MNRAS.456.2749O,1505.04146,2015MNRAS.447..400A,2017MNRAS.464.4938W}. Moreover, significant methodological advances on precursor instruments, such as $m$-mode analysis \cite{1302.0327,Shaw:2014vy}, have been done but more work is needed to apply them to realistic mock data.
    \item Cosmological processing, such as producing sky maps or power spectra.  These steps are quite similar to those routinely undertaken in the CMB and large-scale structure communities, though the fact that we are dealing with interferometric data after aggressive foreground cleaning brings its own unique set of challenges.
\end{itemize}

This pipeline will be essential for testing instrument designs and establishing requirements.  As just one example, a significant systematic effect arises when the instrument couples polarized signals into unpolarized signals - polarization leakage as discussed in Section \ref{sec:optic-mech-antenna}. Only careful computer modeling can enable us to set hardware requirements for such leakage.

One particular issue for PUMA will be deploying known solutions `at scale'.  Radio astronomy has traditionally pushed the boundaries of big data in astronomy and PUMA will be no exception.  \stageone\ $21\,$cm experiments already produce $\mathcal{O}(100\,TB)$ of data per day and PUMA will eventually produce a data set over $100\,\mathrm{PB}$ even assuming compressions exploiting the redundancy of the array.  Further compression through co-addition of daily maps into weekly maps etc.~will be required before the data can be transported and analyzed.  All of this puts strong demands on real-time instrument calibration and characterization.

\section{R\&D prong \#3: Real-Time Signal Processing \& Calibration}
\label{sec:realtime}

The success of PUMA hinges on the success of the real-time calibration and FFT-based correlation techniques. Neither of these techniques has been demonstrated at the precision required for intensity mapping in a working telescope. 

The standard approaches for calibration of intereferometric telescopes with a large number of elements fall into two broad categories. The first class is the so-called sky model calibration, which uses a known catalog of point sources (or other information about the signal in the sky) to measure the complex gains \cite{2016ApJ...833..102B,2016PASA...33...19T,2017MNRAS.464.1146H}. The second class, known as redundant calibration, uses the redundancy of the baselines formed by array elements on a grid to solve for individual complex gains without requiring a sky model; instead it makes strong assumptions about self-similarity of elements and their distribution on the lattice \cite{1992ExA.....2..203W,2010MNRAS.408.1029L}. Both of these classes of methods are imperfect: it is possible that our knowledge of the radio sky is not deep enough to enable sufficient calibration, and also that the elements are not sufficiently self-similar to enable purely redundant calibration \cite{2018ApJ...863..170L}. In addition, PUMA needs to perform calibration not just accurately enough, but also in real time. The methods employed will likely combine both approaches.

The correlator as described in our reference design (see section  \ref{sec:correlator--real}) will contain a dedicated subsystem whose main purpose will be to keep the telescope calibrated and supply frequency dependent complex gain vectors to the FFT correlator.
The purpose of this R\&D prong is to develop and validate algorithms that will enable FFT correlation and real-time calibration and therefore inform the precise design of the \emph{Calibrator} subsystem of the correlator. This effort will be effecively integrative: it will use the tools developed under prong \#2, together with lessons learned about hardware in the laboratory (prong \#1), in the field (prong \#4) and potentially from other experiments to simulate possible inputs to the \emph{Calibrator} and asses the quality of calibration solutions.  With the available tools we should be able to simulate any observable quantity in the presence of realistic complex gain errors, antenna imperfections, geometry imperfections, etc.

In the current instance we have scoped the \emph{Calibrator} subsystem to enable direct correlation for approximately 1\% of baselines, which makes its computational intensity comparable to that of the FFT correlator. In practice, there are three natural choices for which baselines to correlate: i) Correlate 0.5\% of antennae with all the remaining antennae; ii) correlate all antennae for 1\% of the bandwidth and iii) correlate a random subset of baselines. In each case we also need to specify how often we change the subset of baselines or bandwidth we are measuring. There are advantages to each possibility, but the details of the optimal strategy also depend on the achieved spectral smoothness and stability of the instrument. The algorithm will in general use this information, potentially in combination with the FFT correlator output, to compare with sky catalogs of known sources to iteratively correct the phase solution of the entire system. 

Once we learn about the efficiency of the calibration system we can build an effective model that allows us to emulate the phase and gain errors of the output of the FFT correlator as stored on disk. This information will then feed back to the simulation pipelines to allow an accurate simulation of the final on-the-disk data products.

Iterations on the design are expected throughout the course of the R\&D and Final Design phases, with each stage benchmarked against error and systematics budgets. Here we make some preliminary observations:
\begin{itemize}
\item Assuming one minute integration of one antenna with all the rest gives a $10\,\sigma$ point source sensitivity in the mJy range. There will always be point sources of this brightness in the primary beam of the antenna and very often significantly brighter ones.  Questions remain as to whether our knowledge of the sky at the relevant level is sufficient, if the SNR is sufficient and if the presence of other signals that are of scientific interest will bias the calibration.

\item An externally injected pulsating signal (for example, from a central tower in the near-field) can provide precise phase information. The presence of switching allows isolation of the calibration signal and perhaps the fact that the signal enters through deep sidelobes does not matter.

\item It might also be possible to consider naturally pulsating celestial sources such as pulsars, which are effectively point sources and relatively large in number at PUMA sensitivities. It remains to be seen if obtainable SNRs will be sufficient.
\end{itemize}

We note that there exist algorithms (such as EPICal \cite{1603.02126}) for self-calibration of the telescope that do not require any brute-force correlation. These algorithms have not been demonstrated to work at the required accuracy for PUMA, but offer an alternative path. It is possible that the PUMA Calibration scheme would use several of these techniques in unison.

\section{R\&D prong \#4: Path-finder arrays}
\label{sec:pathfinders}

\subsection{PUP engineering prototypes}

The final pillar of our development path are path-finder arrays. Their main purpose is to validate the hardware designs and computer simulations described above. These pathfinder arrays are unlikely to have the sensitivity to make an actual measurement of the cosmic signal, instead they should be thought of as engineering prototypes in preparation for the final design.  We propose a four-stage effort with small telescopes referred to as PUPs (PUma Prototypes), with increasing scope of complexity and validation. These prototype arrays will deploy the same on-the-focus digitization hardware and clock distribution, but will employ direct $N^2$ correlations so as to have full understanding of system properties.

In Table \ref{tab:pups} we discuss a possible progression of these prototypes, associated timelines and the goals of this validation exercise. The described program can start soon and should last well into the design and fabrication phase of the final array in order to have a continuously working testbed. When expanding the dishes, especially in the early stages (e.g. from PUP-2 to PUP-6, or from PUP-6 to PUP-20) we might replace all of the dishes if the design evolves as dictated by test results from the previous prototype. If not, the expanded array will be built as an expansion of a smaller pathfinder. Based on our experience with the BMX testbed at BNL, we think that these prototypes should be located for convenient access rather than in an exclusive radio-quiet zone.

We conclude that despite the tremendous value of working with real hardware, some of the anticipated challenges of the full-array will remain inaccessible to the pathfinder arrays. It is unlikely that we would be able to validate a full real-time calibration scheme due to lack of total signal-to-noise and would be unlikely to validate our wedge cleaning algorithms, due to insufficient density of baselines. Nevertheless, in combination with our numerical simulations, we believe we can validate our approaches with a degree of certainty.

\begin{table}
  \centering
  \begin{tabular}{c|c|c|c}
    Name & Description & Purpose & Timeline \\
    \hline
    PUP-2 & \begin{minipage}{4cm}
      A pair of 6m dishes on a $\sim$30\,m E-W baseline
      \end{minipage}   & \begin{minipage}{8cm}
      \
      \begin{itemize}
      \item Perform the basic checks of the system, obtain first fringes and auto-correlation transits

      \item Validate antenna design

      \item Independently measure and characterise the two beams

      \item Measure beam stability as function of pointing, temperature, humidity, etc
      \end{itemize}
      \
      \end{minipage}
      & 2020-2022 \\ \hline
    PUP-6 &\begin{minipage}{4cm}
            Six dishes in a packed cluster at PUMA configuration
     \end{minipage} & \begin{minipage}{8cm}
      \
      \begin{itemize}
      \item Measure dish cross-talk and shadowing

      \item Measure dish-to-dish variations and manufacturing tolerances

      \item Validate pointing tolerances

      \end{itemize}
      \
      \end{minipage}
      & 2022-2023 \\ \hline

      PUP-20 & \begin{minipage}{4cm}
          20 dishes in a packed cluster at PUMA configuration
          \end{minipage} & \begin{minipage}{8cm}
      \
      \begin{itemize}

      \item Measure cross-talk and coupling between dishes in the ``continuum limit''

      \item Continue measuring beam variations, tolerances with larger sample

      \item Validate final model for components that describe dish-to-dish beam variations

      \end{itemize}
      \
      \end{minipage}
      & 2023-2024 \\ \hline

      PUP-60 & \begin{minipage}{4cm}
        Three clusters of 20 dishes in a packed cluster at PUMA configuration separated by $\sim$\,km baseline
        \end{minipage} & \begin{minipage}{8cm}
      \
      \begin{itemize}
      \item Validate phase calibration techniques

      \item Measure and validate phase stability on long baselines

      \item Measure the effects of gradients of environmental properties across the array

      \end{itemize}
      \
      \end{minipage}
      & 2024-2028 \\ \hline

  \end{tabular}
  \caption{Proposed pathfinder arrays to be built as engineering prototypes for demonstration and validation of PUMA technology. The number of dishes specified here are indicative -- the main purpose is to have a testbed of both closed-packing of antennae to asses the level and all sources of coupling and the sufficiently long baselines to asses the phase stability.}
  \label{tab:pups}
\end{table}

\subsection{International Context}

Over the last decade, interest in post-reionization HI intensity mapping has grown and several \stageone\ experiments are underway or proposed. Three of them (TianLai, HIRAX, CHORD) include close-packed interferometric dish arrays that are conceptually similar to PUMA.
Currently US participation in these projects is minimal. A small-scale, coordinated US investment in one or more of these experiments could accelerate progress in the field, especially if coordinated with the more long-term R\&D towards PUMA.

\section{Reference Design}

PUMA is a proposal in a pre-conceptual stage and the sections so-far have argued that we have a clear path for R\&D that will lead to possible realization of this experiment. We now discuss our reference design and how we envision all the pieces working together. Although the system and its specifications is defined in the outline, many details are not currently known. The main purpose of this reference design is to allow us to cost the project with reasonable accuracy and to compare various trade-offs entailed in detailed design decisions.

\subsection{Antenna Array and Survey Overview}

The basic parameters of our reference array design are fixed by scientific considerations and were discussed earlier in the text. The main array is composed of $6\,$m dishes occupying lattice points of hexagonal close-packing with 50\% fill factor. We show one example of such packing on the title page. The primary amplification and digitization of the signal is done at the dish and the signal is routed thereafter on optical fiber cables. A $100\,\mathrm{Gbit/s}$ optical fiber cable is capable of carrying a full-band unchannelized signal from both polarizations of a single antenna.
Channelization and band selection is done in small units covering $\sim$6 dishes each, that sit on the unoccupied lattice sites.  Once the datastream is channelized and unused spectral bands discarded, a single fiber can carry the signal from a few elements, so the input to the correlator will be several thousand optical fibers.

As with any modern survey experiment, the survey and experiment are tightly coupled. While with any transit telescope there is no freedom to track, we do have freedom to re-point in declination thus affecting the trade-offs between depth and observed area. Our science goals, plus the need to avoid observing close to the horizon, have led us to a fixed depth survey with 50\% of the extragalactic sky covered.

The main properties of the Antenna Array and Survey are summarized in Table \ref{tab:array}. The potential sites are discussed in Section \ref{sec:site-select-observ}.

\begin{table}
  \centering
		\begin{tabular}{|l|C{4cm}C{4cm}|}
			\hline
                  Antenna Array & & \\
                  \hline
                  \quad Element Distribution  & \multicolumn{2}{c|}{Hexagonal close-packed transit array} \\
                  \quad Lattice Spacing & \multicolumn{2}{c|}{6m} \\
                  \quad Lattice Fill-factor & \multicolumn{2}{c|}{50\% random lattice sites} \\
                   && \\
                                & {\bf PUMA-5K} & {\bf PUMA-32K}  \\
			\quad Array Diameter & 630m & 1500m  \\
			\quad Number of Elements & 5,000 & 32,000  \\
                        \quad Angular Resolution & 1.5' - 8' & 0.6' - 3.2'\\
                        \quad $10\,\sigma$ Single Transit Sensitivity (for source at decl. $-30^\circ$) &     8.7$\mu{\rm Jy}$ & 1.3$\mu{\rm Jy}$  \\
			\hline

                  Survey  & & \\
                  \hline
			\quad Observing Area & \multicolumn{2}{c|}{50\% of the extragalactic sky (see also Sec. \ref{sec:site-select-observ})} \\
                        \quad Observing Time & \multicolumn{2}{c|}{5 years on sky, wall-time 7-10 years} \\
                             &&\\
                                & {\bf PUMA-5K} & {\bf PUMA-32K} \\
			\quad Equivalent Source Density ($z=2$, $k=\SI{0.2}{\hPerMpc}$) & $\SI{2.0}{\h\tothe3\per\Mpc\tothe3}$  &  $\SI{7.4}{\h\tothe3\per\Mpc\tothe3}$  \\
			\quad Total Equivalent Sources  & & \\
			\quad \quad at $k=\SI{0.2}{\hPerMpc}$ & 0.6  billion & 2.9 billion\\
			\quad \quad at $k=\SI{0.5}{\hPerMpc}$ & 0.4 billion & 2.5  billion \\
			\quad FRB rates (expected)  & &\\
			\quad \quad  200-400\,MHz (\emph{highly uncertain})& 70/day & 1200/day \\
			\quad \quad  400-700\,MHz & 60/day & 1000/day \\
			\quad \quad  700-1100\,MHz & 80/day & 1300/day \\
			\hline

                \end{tabular}
		\caption{\label{tab:array} Basic parameters of the array and survey. See also Tables \ref{tab:antenna}, \ref{tab:corr} }
	\end{table}

\subsection{Antenna Element Overview}

\begin{figure}
    \includegraphics[width=\linewidth]{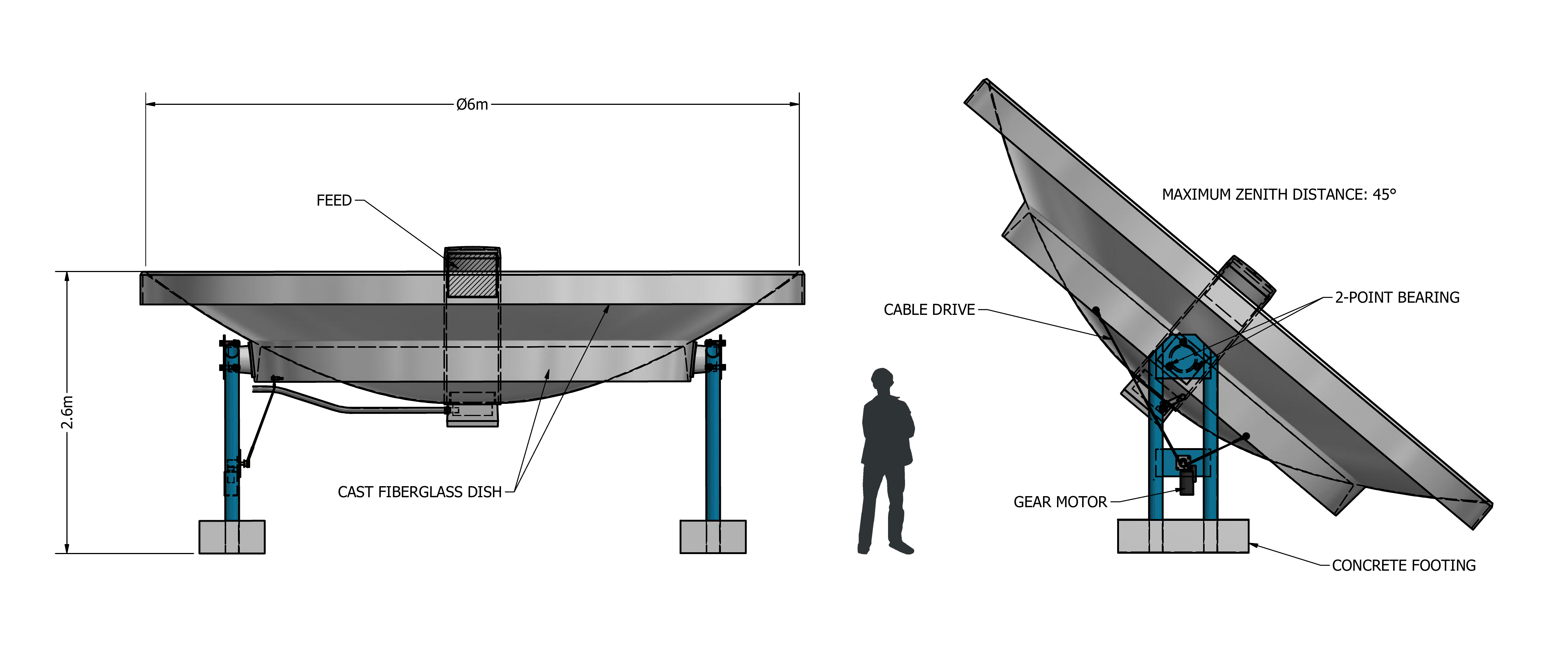}
    \caption{Notional dish and mount design. Moving mass $\sim$ 500kg.}
    \label{fig:dishGraphic}
\end{figure}

We show a conceptual model of a PUMA $6\,$m dish with simple mount steerable in altitude in Fig.~\ref{fig:dishGraphic}. It incorporates a 7mm-thick molded paraboloid reflector reinforced by a co-molded support ring, supported by altitude bearings at $\sim$1.8m above grade along an E-W baseline; this allows the dish to be tilted up to 40$^{\circ}$ off-zenith, with the center of gravity positioned to avoid angle-varying load on the supports. Dish construction methods are discussed in Sec.~\ref{sec:dishConstruction}. Notionally, the wideband feed and (uncooled) receiver electronics will be located within a cylindrical pillar supported from the dish vertex -- this has several advantages over the more standard approach employing spider legs, notably in terms of cabling of the output signal and weather protection; Cassegrain or offset feed geometries will also be modeled. Studies to determine the optimal $f/D$ ratio considering sensitivity, degree of dish illumination, and edge taper have not yet been done. We estimate \cite{Stage2WhitePaper} an overall aperture efficiency ${\eta_a}\simeq 0.7$ at $1\,$GHz.

\begin{table}

  \begin{tabular}{|c|c|c|}
\hline
    \textbf{Antenna Array Element} & \textbf{Value} &\textbf{ Comment}\\
 \hline
Sizes of Array Elements or Segments& $\Phi 6$ m & parabolic dish\\\hline
Number of Array Elements or Segments&  - & see Table \ref{tab:array}\\\hline
Field of View& 3.1 - 17.2$^{\circ}$ FWHM&(primary beam)  \\\hline
Wavelength range& 0.27 - 1.49 m& \\\hline
Optical surface figure quality (RMS)& $\sim$ 1mm&TBD \\\hline
Uniform separation between Array Elements& - &see Sec. \ref{sec:shadowing} \\\hline
Mass of each Optical Element & 500 kg& \\\hline
Degrees of Freedom& 1&altitude \\\hline
Type of mount used for pointing and allowed range&0$^{\circ}$ - 40$^{\circ}$&from zenith\\\hline
Mass and Type of Material for Support Structure &- &see Fig. \ref{fig:dishGraphic} \\\hline
Optic Design (e.g., Cassegrain)& prime focus feed &see Fig. \ref{fig:dishGraphic} \\\hline
Description of Adaptive Optics& none& \\\hline
Spectral Range &200 - 1100 MHz &  \\\hline
Number of detectors, type, and pixel count & 1&dual-polarization feed  \\\hline
Thermal or Cryogenic Requirements & - &stabilized receiver enclosure \\\hline
Size/dimensions (for each instrument) & 10$\times$10$\times$10 & m$\times$m$\times$m \\\hline
Instrument average science data volume per day/night  & - & see Table \ref{tab:corr} \\\hline
Development Schedule & 50 - 80& months \\\hline
  \end{tabular}
  \caption{Telescope/Antenna Array Characteristics Table. \label{tab:antenna}}

\end{table}


\subsection{Correlator \& Real-time Calibration}
\label{sec:correlator--real}

\begin{figure}
\centering
\includegraphics[width=0.90 \linewidth, clip]{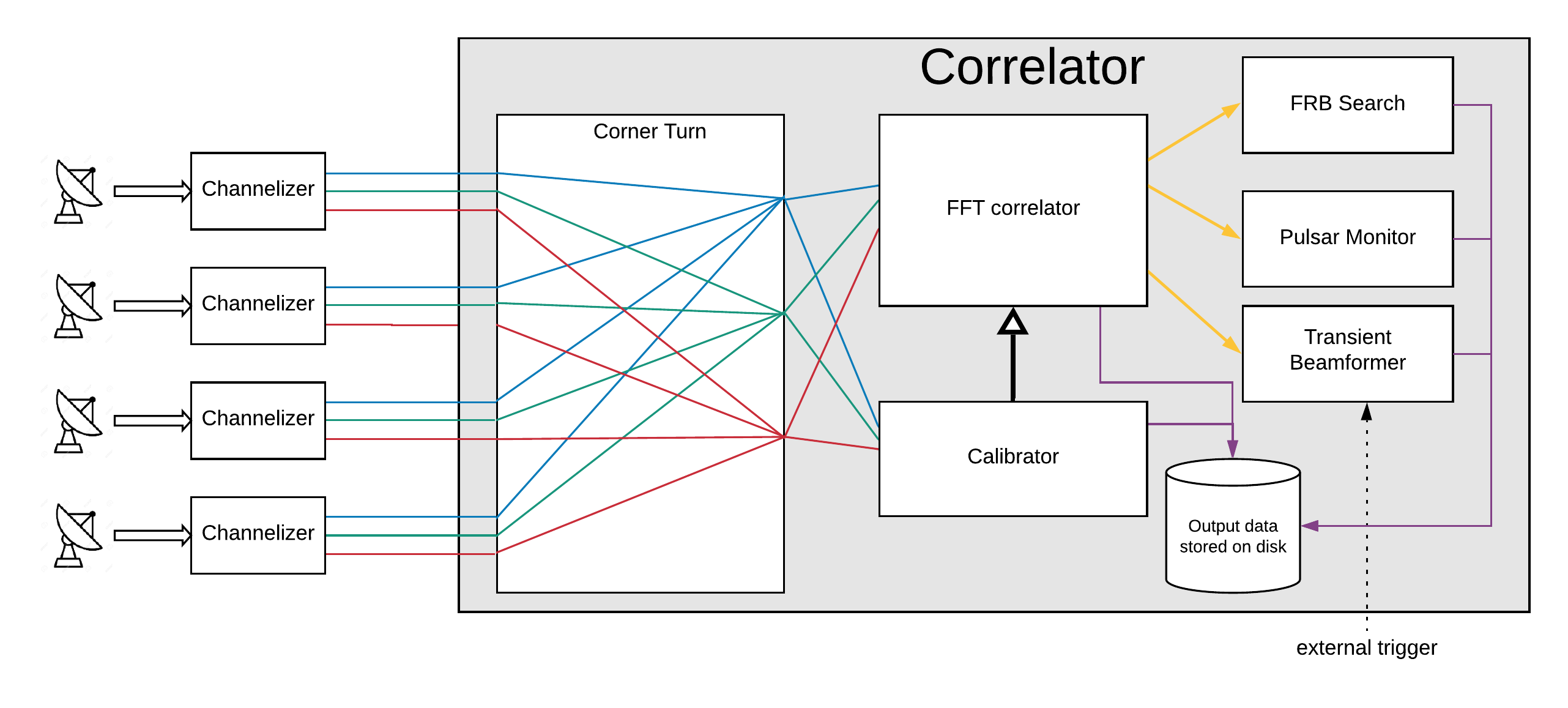}
\caption{Abstract conceptual diagram of the required data transport in
  a wide-band mapping array.  The digitization of the signal takes place in the antenna feed housing, followed by channelizer in the vicinity of the antenna to lower the bitrate. The channelized signal -- different colored lines represent different frequency bins -- is then transported to the centrally hosted correlator where it is processed as indicated schematically (see Sec.~\ref{sec:correlator--real}). }
\label{fig:data_transport_abstract}
\end{figure}

The PUMA correlator is composed of six main subsystems. The overall data path is indicated in Fig.~\ref{fig:data_transport_abstract} and summarised in Table \ref{tab:corr}.
Appendices \ref{app:ccost} and \ref{app:drates} outline our methodology for computing computational intensity and data rates.  The subsystems are:

\subsubsection*{Corner-turn}

After digitization and Fourier transform of the RF waveforms
at the receivers (see Section~\ref{sec:integratedFrontEnd}), the channelized data with all the frequencies from a single antenna are present at one point.
The correlation step, on the other hand, expects the data from all the antennae for a given frequency channel to be
available together.  Repackaging the data is known as a ``corner-turn'' problem. In our implementation the corner-turn is a part of the correlator which
interfaces the antenna outputs to the actual correlator input. This is done straightforwardly in a layered network containing
$\log_2 N_{\rm ch}$ layers (where $N_{\rm ch}$ is the number of spectral channels).

\subsubsection*{FFT correlator}
\label{subsec:FFT_correlator}

At the size of the PUMA array, the FFT based correlation is the only realistic choice to extract the (almost) complete information in the datastream. The FFT correlator will implement the FFT correlation algorithm  and dynamically co-add  all the redundant baselines. This leads to massive data compression factors, at the cost of stringent demands on real-time calibration. At the same time, the FFT correlator will also stream beam-formed images of the current sky to the FRB search engine, Pulsar monitoring engine and Transient Beamformer for further processing.

\begin{table}
  \centering
  \scriptsize
  \begin{tabular}{c|p{5cm}|c|c|c}
    Subsystem & Description & Scaling &   PUMA-5K & PUMA-32K \\
\hline
    Corner-turn & Rearranges data from antenna orderd to channel ordered & $N\times N_{\rm ch}$ & 180Gbs throughput & 1.2 Pbs throughput \\
     \hline
    FFT correlator & Full field redundant baseline co-addition employing FFT algorithm & $N\log_2 N$ &
\begin{minipage}{3cm}\ \\\begin{centering}$\sim$15 PFLOP/s\\  $\sim$ 1 PB/day\\ \end{centering}\end{minipage} &
\begin{minipage}{3cm}\ \\\begin{centering} $\sim$ 100PFLOP/s \\ $\sim$ 7 PB/day\\ \end{centering}\end{minipage} \\
    \hline

  Calibrator & Direct correlation for $\sim$ 1\% of baselines for real-time calibration, storing a small fraction for off-line QA & $N^2$ &
\begin{minipage}{3cm}\ \\\begin{centering} $\sim$ 2 PFLOP/s  \\ Saving as required for QA \end{centering}\end{minipage} &
\begin{minipage}{3cm}\ \\ \begin{centering}$\sim$ 70 PFLOPS/s \\ Saving as required for QA \\ \end{centering}\end{minipage} \\
    \hline
    FRB Search engine & Real-time search for Fast Radio bursts (uses FFT correlator output as input)  & +10\%  &   \\
    \hline
    Pulsar Monitoring & Monitoring of in-beam pulsars (uses FFT correlator output as input)  & +10\%   &   \\
    \hline
    Transient Beamformer & Storing fine time or frequency resolution data based on external triggers (uses FFT correlator output as input)  & +10\%   &   \\

  \end{tabular}
  \caption{Properties and subsystems of the PUMA correlator.}
  \label{tab:corr}
\end{table}

\subsubsection*{Calibrator}

This subsystem will continously update the FFT correlator with complex gains solutions. Its development is a major effort that we describe in Section \ref{sec:realtime}. The system has been baselined to allow 1\% of all baselines to be correlated by brute force.  This number has been set by what we believe is reasonable rather than by quantitative study, which is the goal of prong \#3. This still produces enormous data rates, especially for the large baselines, so this information will not be stored, except for a small fraction of summary quantities to monitor the health of the overall system offline. Instead this information will be used for real-time calibration and discarded.

\subsubsection*{Other Correlator Engines}
There will be three more engines that will feed from the full output of the FFT correlator. Since the FFT correlator will be fed by the polyphase filter-bank (PFB) samples, the complex outputs of the FFT correlator will allow many of the usual transformations, including increasing or decreasing the native PFB spectral resolution and various slicing and integrations in time and frequency. The three additional engines are low risk, since the calibration requirements on these are considerably lower and they are considered solved problems.
We consider three additional engines:

\minisection{FRB Search Engine} will look for new FRBs using algorithms that have been successfuly applied to the current generation of experiments \cite{1702.04728,1803.11235} and will be further developed over the coming decade.

\minisection{Pulsar monitoring} will form beams at the positions of known pulsars and integrate and measure them as they cross the sky.

\minisection{Transient Beamformer} will be a target of opportunity engine that will save as much information as possible on some beams, as triggered by externeal observatories such as gravitational wave detectors and perhaps nearby supernovae explosions (there is $\sim$ 10\% chance that a supernova will go off in our galaxy over ten years of observations with PUMA).

\subsection{Other considerations}

In this section we discuss a few other parts of the design being considered for PUMA.

\subsubsection*{Antenna Shadowing}
\label{sec:shadowing}
Throughout this development, we have considered a perfectly close-packed array, while at the same time assuming an observed fraction of the sky that requires at least some repointing. This will necessarily entail some shadowing of the nearby antennae. This can be avoided by stretching the array configuration appropriately in the N-S direction, but the changes are relatively modest. To avoid purely geometrical shadowing one needs to stretch by $\sim 15\%$ for hexagonal packing and $\sim 10\%$ for an optimally rotated, square close packing  (the exact numbers depend on the maximum offset from zenith, which depends on the observatory latitute). In practice, these numbers will likely need to be increased a little, but these corrections are within the errors in our forecast.

\subsubsection*{Lattice offset for antenna positioning}
An interesting approach towards distribution of array elements has been taken by HERA, whose antennae lie on a hexagonal lattice, but one where the array is divided into three sub-sections with subsections having half-lattice spacing offsets. This allows baselines that cover distances in the $u-v$ plane that are not otherwise possible for perfect hexagonal packing. This leads to a more uniformly filled $u-v$ plane, but also increased demands on the FFT correlator by a factor of $\sim 4$. PUMA did not take this approach in the version presented in this document. However, if the simulation results of our calibration scheme (described in the Section \ref{sec:realtime}) indicate that this is needed for successful foreground control, we might incorporate this into our design.

\subsubsection*{Alternative Pointing Strategies}
The current design of PUMA relies on array elements that can change in elevation in the N-S direction, but are not capable of tracking celestial sources. We subscribe to the current thinking in the field that posits that beam stability is more important than tracking and it is most easily obtainable with a stationary telescope. Thus our observing strategy is based around occasional repointing and recalibration of the system, but relying on Earth rotation for sky scanning.

An alternative to the elevation-only strategy is to instead adopt a mounting scheme in which dishes are at fixed zenith angle, but are able to rotate in azimuth. This design has some advantages. It still allows a range of declinations to be observed (i.e.\ when pointing purely east/west, it would see the same sky as if pointing to the zenith, although somewhat earlier), while keeping the gravity vector affecting the dish (and thus primary beam to a large extent) constant. It could potentially allow a continuous rotation in azimuth during observations which would modulate the sky signal in a unique way. It would allow observing the same declination at two values of azimuth, which would correspond to rotation of our OMT with respect to the sky and thus modulate polarization response. We did not adopt this solution in the reference design as it seems mechanically cumbersome, but we are keeping our options open as we learn more about design trade-offs.

\subsubsection*{FRB Outrigger Stations}
The current generation of FRB search engines contain remote stations that can be used for precise localization of the FRBs once they are detected. This system works by triggering a ringbuffer dump at all participating telescopes once an FRB is detected and cross-correlating the signal off-line in a way that is similar to VLBI.
The reference PUMA design does not contain any such station. On one hand, the native resolution of PUMA will be a factor of a few bigger than that of current experiments (due to its larger size) and on the other hand the next decade of FRB research will show just how useful this localization is. We will reconsider this once the project reaches further maturity and in the context of concurrently-running experiments.

\subsubsection*{Real-time expansion of the Universe.}
One of the interesting potential science application of 21\,cm observations is direct measurement of the cosmic expansion by measuring changes in redshift in cold 21\,cm absorption systems (see \cite{1311.2363,Jiao:2019alc}, but also the discussion in \cite{Stage2WhitePaper}). This is considerably easier in radio than in optical, because ultra-precise atomic clocks that are accurate at $10^{-10}$ level can be obtained off-shelf. Because we need exquisitely stable clock distribution for the phase calibration, the only real requirement in terms of hardware is to use a sufficiently precise atomic master clock and an additional subsystem in the correlator that is capable of measuring the objects of interest with the very high spectral resolution. Pursuit of this science case has recently begun within the CHIME collaboration; we have not developed it in detail for PUMA, but it would be worth revisiting this problem as the project matures.

Similarly, PUMA may be able to measure the Hubble constant using strongly lensed FRBs through ultra-precise time-delay cosmography. In contrast to traditional time-delay cosmography using quasars, which is sensitive to the lens model \cite{2019arXiv190704869W}, with FRBs one can track the increase of the lensing distance with the expansion of the Universe, providing a measurement of the Hubble constant without needing to model the lens \cite{2018ApJ...866..101Z}.  Roughly one out of every thousand cosmological lines of sight are strongly lensed by a galaxy so the PUMA FRB sample will contain of order one thousand strongly lensed FRBs which can be identified in the time domain from stellar microlensing \cite{2019arXiv190706830O}. Furthermore, a significant fraction of FRBs are seen to repeat \cite{2019ApJ...885L..24C} and so PUMA will likely discover several strongly lensed repeaters.  These can then be monitored to detect the delayed copies of each burst from each lensing image. The flashing nature of FRBs allows the delays to be measured to exquisite precision of $10^{-10}$. Tracking the change of the delays for a repeating FRB over far-separated epochs allows one to measure the change in the lensing distance due to the Hubble expansion and thus the Hubble constant. PUMA, with nominal capabilities, will discover these strongly lensed repeaters and the outriggers described above would facilitate followup. Observing the arrival of each burst image requires high-duty-cycle monitoring, and thus followup from other instruments.

\section{Facilities, Operations \& Observational Strategy}

Facilities, Operations and Observational Strategy are the least developed parts of our plan towards PUMA realization. This is driven by the fact that these are not crucial aspects of R\&D, but rather can be developed once the experiment reaches a more advanced stage.

\subsection{Site Selection and Observational Strategy}
\label{sec:site-select-observ}

Remote southern-hemisphere sites, which are known to have low levels of RFI and which have already-built infrastructure for large radio observatories, are favored for PUMA. Such sites would also have maximum sky overlap with major new facilities at other wavelengths, providing opportunities for cross-correlation science (these science cases were studied in some detail in the \stagetwo\  whitepaper \cite{Stage2WhitePaper}):

\begin{itemize}
\item \textbf{LSST} is the main optical survey of the southern sky. It will measure 4 billion galaxies to redshift $z\sim3$. Cross-correlations with PUMA can potentially enable photometric redshift calibration (assuming long modes parallel to the line of sight are reconstructed), thus greatly improving key LSST science.

\item The upcoming \textbf{Simons' Observatory} and the proposed \textbf{CMB-S4} are major CMB experiments that will cover significant fractions of the southern sky from the Atacama Desert in Chile and the South Pole. By studying deformations in the CMB due to gravitational lensing they will reconstruct the projected matter density field over an enormous volume. Sky overlap between the next generation CMB experiments and PUMA will enable exciting tests of our cosmological models.

\item \textbf{SKA} will be a synergistic radio telescope that will survey the southern sky at high resolution and across frequencies. It will provide the pulsar sample for PUMA as well as (potentially) a catalog of calibration sources.
\end{itemize}

Siting PUMA in the southern hemisphere will also increase the number of accessible pulsars (since the galactic center is in the south). Finally, together with the Canada-led program of FRB discovery engines (CHIME and the proposed CHORD experiment) and the proposed DSA-2000 instrument \cite{hallinan2019dsa}, both of which are in the northern hemisphere, a southern PUMA would complete the census of repeater FRB sources across the observable sky.

Clearly, a formal site selection process that takes into account PUMA's observing requirements, cost considerations, and inter-agency and international agreements will need to be developed. With its modest angular resolution requirement, PUMA's physical footprint will be small compared to planned projects in South Africa and western Australia and the opportunity to share infrastructure costs could make a co-siting arrangement attractive. However, sites with favorable conditions in other regions are not ruled out at this early stage.

\begin{figure}
  \centering
  \includegraphics[width=0.5\linewidth]{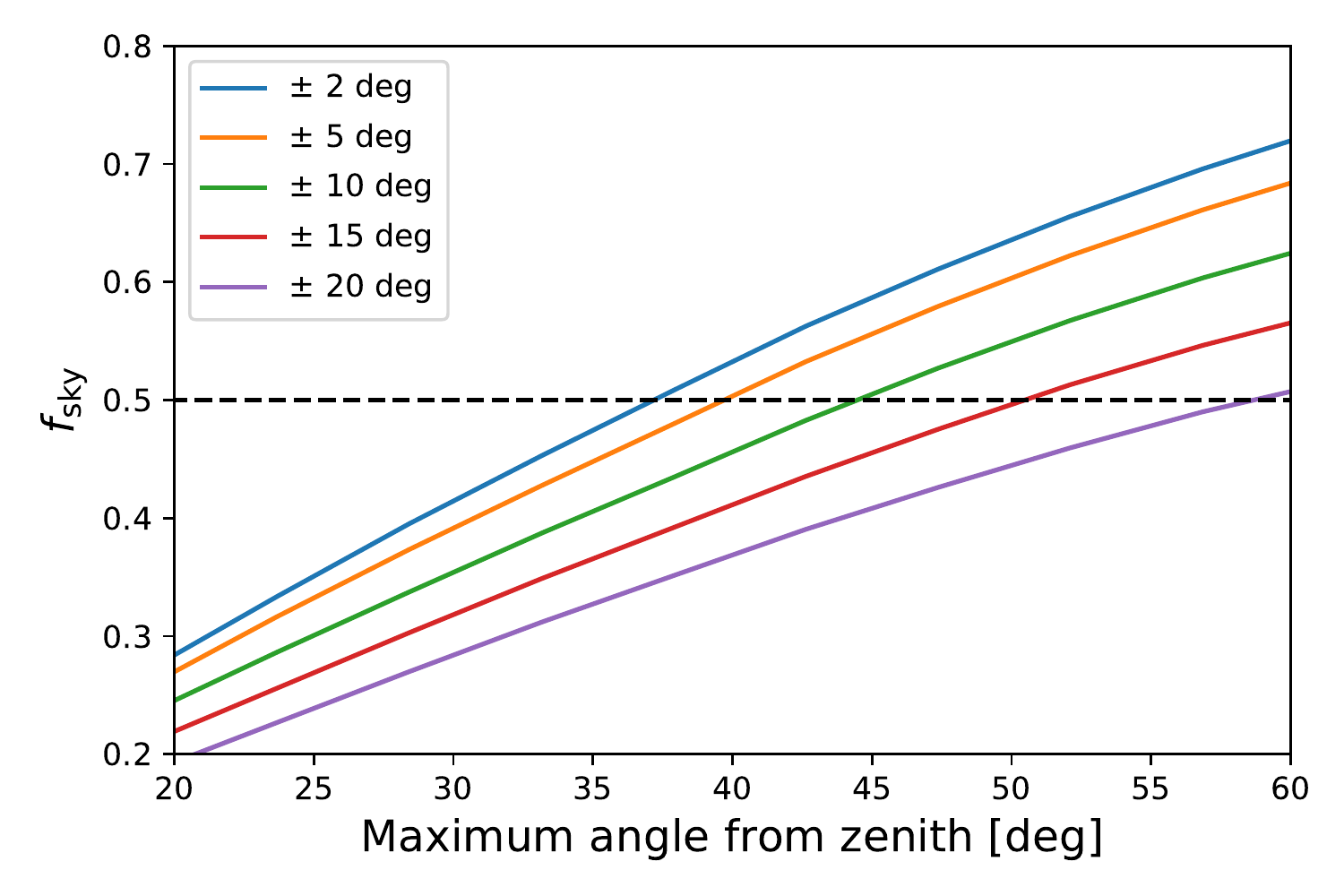}
  \caption{The fraction of accessible celestial sky from an observational site at SKA latitute (30.75$^\circ$S) as a function of galactic cut (lines of different colors as indicated in legend) and maximum angle from zenith. }
  \label{fig:footprint}
\end{figure}

In Fig.~\ref{fig:footprint} we show the fraction of accessible sky from a site at the latitude of SKA in South Africa as a function of maximum observation angle from zenith and galactic cut. We see that with a $40^\circ$ cut we can observe roughly 50\% of the accessible extragalactic sky. The primary beam width increases the accessible area to closer to 60\% at low frequencies.  We will thus cover declinations from $\sim -70^\circ$ to $\sim +10^\circ$, which is ideal for co-observations with LSST.

\subsection{Facilities}

Successful operation of PUMA will require several facilities to be present at the observational site:

\begin{itemize}
\item \textbf{Connectivity.} In order to transfer data, we require connectivity with sufficient bandwidth to support access to real-time health monitoring of the facility and periodic transfer of maps and metadata to a North American archive site. Costs to establish these connections are included in the site upgrade section of the cost estimate. If PUMA is co-sited with an existing observatory, the additional infrastructure will be minimal.

\item \textbf{Electricty.} Total power required for PUMA is estimated assuming the on-antenna receivers will consume 22W each (based on actual analog power for BMX receivers and using the Xilinx power estimator for the on-antenna digital processor). Power to the receivers will be distributed at 48V to the aggregator huts where switching regulators make the conversion to 5/12V at 85\% efficiency.  Power for the non-tracking dish mounts is negligible. Based on CHIME and DSA-2000 estimates, and accounting for reductions of 25\% with technology advances, we consider that the PUMA correlator will require 100kW (PUMA-5K) or 700kW (PUMA-32K). Total power is then 230kW for PUMA-5K or 1.5MW for PUMA-32K, and the cost to install a dedicated diesel generator of this capacity is included in the site upgrade section of the cost estimate.

\item \textbf{Dish construction and characterization facility.} At the scale of PUMA, it is very likely that the most economical way to produce large parts is on site. We therefore anticipate a fiberglass  dish construction facility. Part of this facility will also be profilometry of dishes for quality assurance. This facility is costed in the next section.

\item \textbf{Main correlator and service buildings.} Conventional buildings and roads are part of the site upgrade. The building housing the correlator requires adequate RF shielding and HVAC facilities; these have been included in the correlator section of the cost estimate.

\end{itemize}

\section{Programmatic Issues \& Schedule}
\label{sec:programmatic-issues}

The PUMA concept arose within the 21\,cm working group of the Cosmic Visions Dark Energy process under the auspices of the Office of Science, High Energy Physics program of the Department of Energy (DOE). However,  the scale of the project is such that PUMA will likely be a collaboration between funding agencies. Successful joint projects like US ATLAS, US CMS, or LSST may provide a model for multi-agency funding.

DOE and NSF have collaborated on numerous projects employing a variety of models of collaboration. At this stage, it is difficult to predict what would be the most appropriate model for PUMA. Based on previous and current dark energy experiments and their funding models, we expect that it might be possible that the PUMA-5K could be a DOE-led experiment scoped exclusively for the dark energy and early universe science (science goals A-D).  In this case, NSF could join as a partner to fund the science analysis and instrument features that will enable the FRB and pulsars analysis and to provide radio astronomy expertise. 

The scale of PUMA-32K is such that DOE and NSF will both have to contribute as major participants and in addition funding from international collaborators becomes more important. This case would approximately follow the Vera Rubin Observatory LSST model.  In both cases, we expect considerable contribution from international partners, in line with the current generation of survey facilities.

All agencies develop major experiments and facilities according to well-defined processes starting with definition of mission need, driven by input from the community, followed by R\&D leading to formal design reviews before construction funding is allocated. Nevertheless, there are important differences. While NSF is mostly proposal driven, the DOE is project-driven, where projects are adopted when they follow the general scientific program of the agency. Once underway, experiment construction and operations receive guidance and oversight from agency program managers. In our case,  any project development funding is only expected after successful endorsements of both the Decadal Survey of Astronomy and Astrophysics (ASTRO2020) and the Snowmass / Particle Physics Project Prioritization Panel (P5) process of DOE-OHEP. This limits any short term funding over the next few years to generic R\&D and we have adapted our plans to follow this model.

\begin{figure}
    \includegraphics[width=\linewidth]{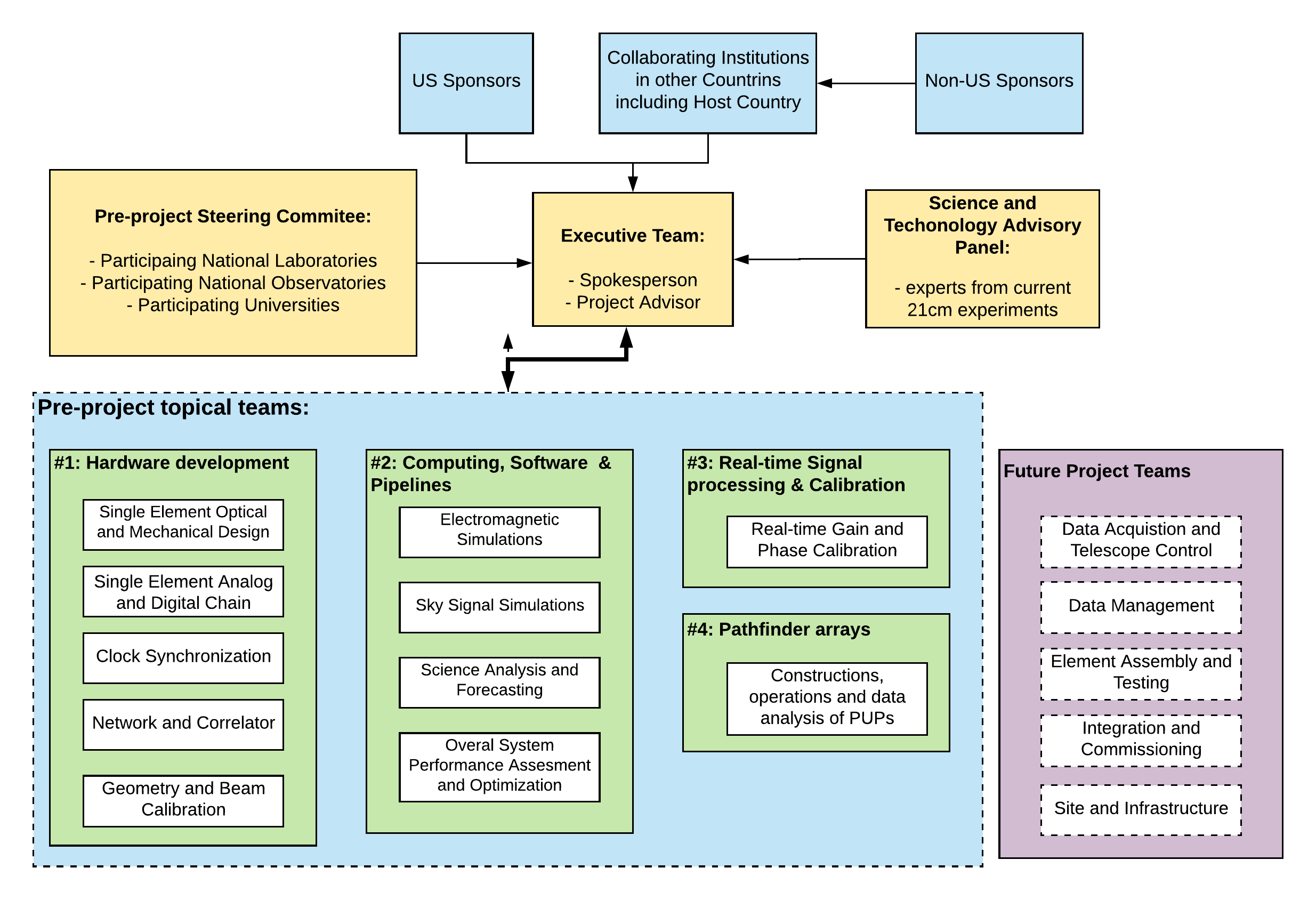}
    \caption{Organizational chart of the pre-project PUMA collaboration. The organization is intentionally loose with weak project and science collaboration separation, which we believe is beneficial at this stage. The organization primary purpose is to act as a vehicle for pre-project funding, enable development and demonstration of necessary technologies and act as embryonic project office and science collaboration.  }
    \label{fig:orgchart}
  \end{figure}

\begin{figure}
  \centering
  \begin{tabular}{cc}
    \multicolumn{2}{c}{\includegraphics[width=0.7\linewidth]{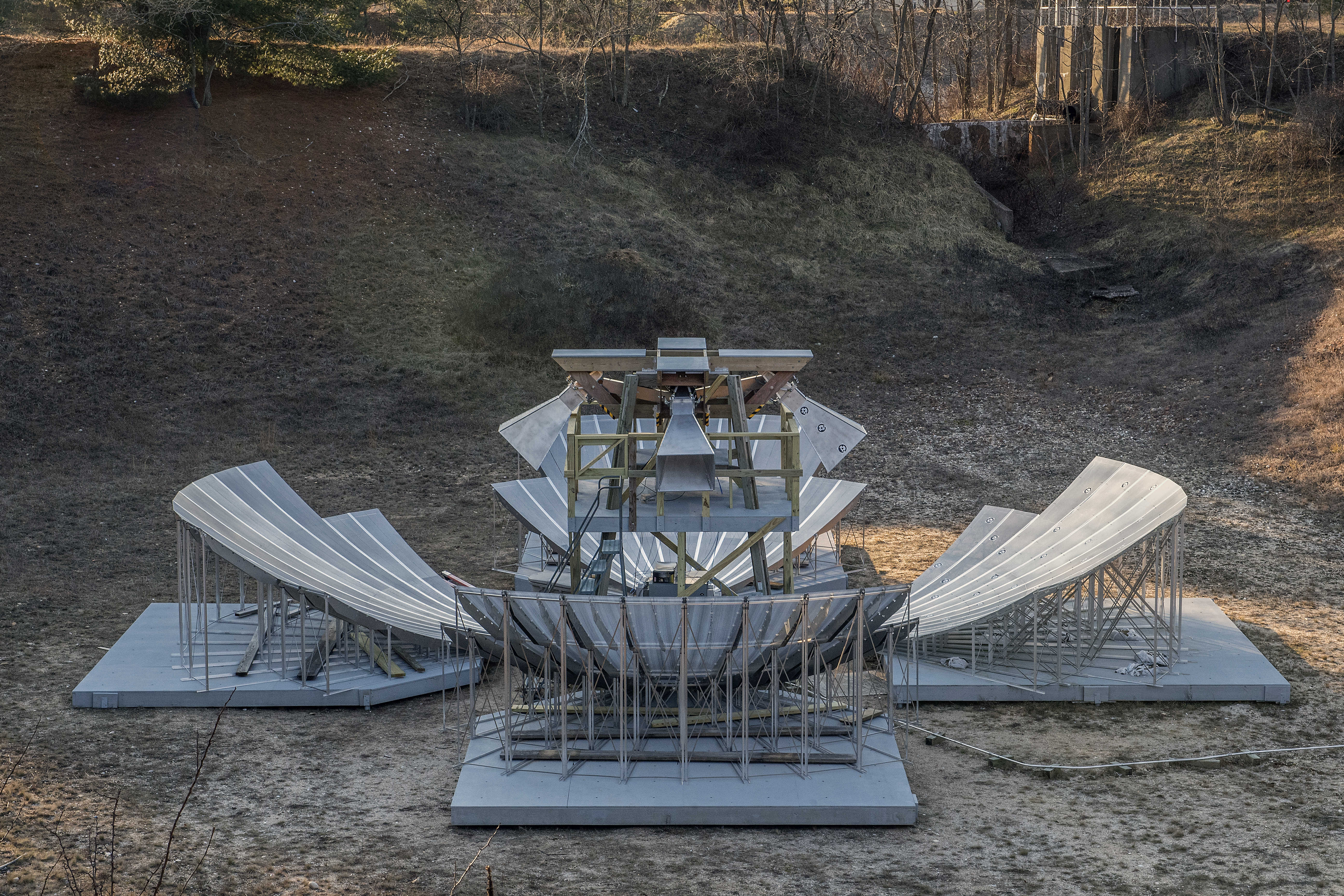}} \\
    \includegraphics[height=0.25\linewidth]{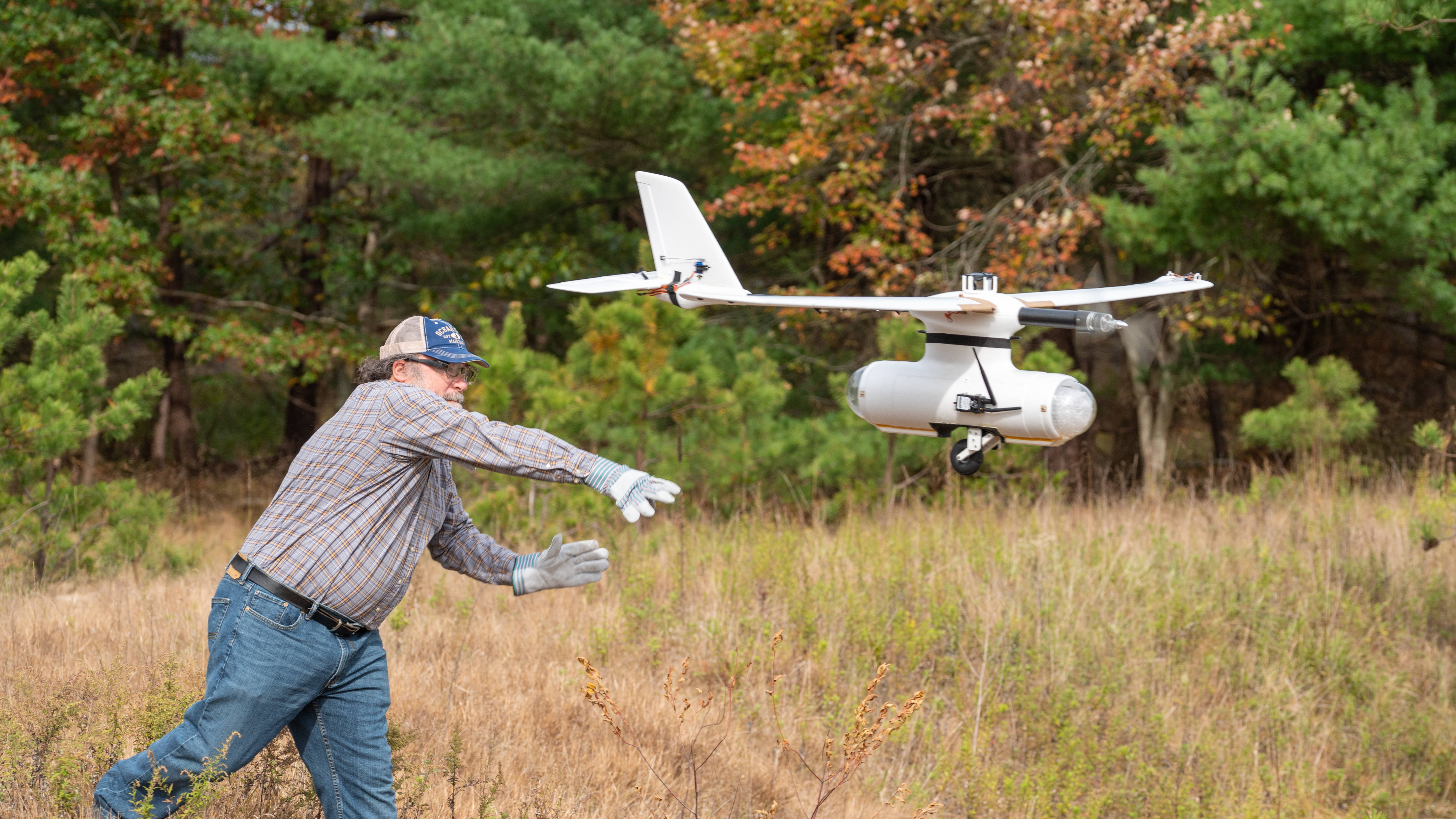} &
    \includegraphics[height=0.25\linewidth,trim=20 30 30 30, clip]{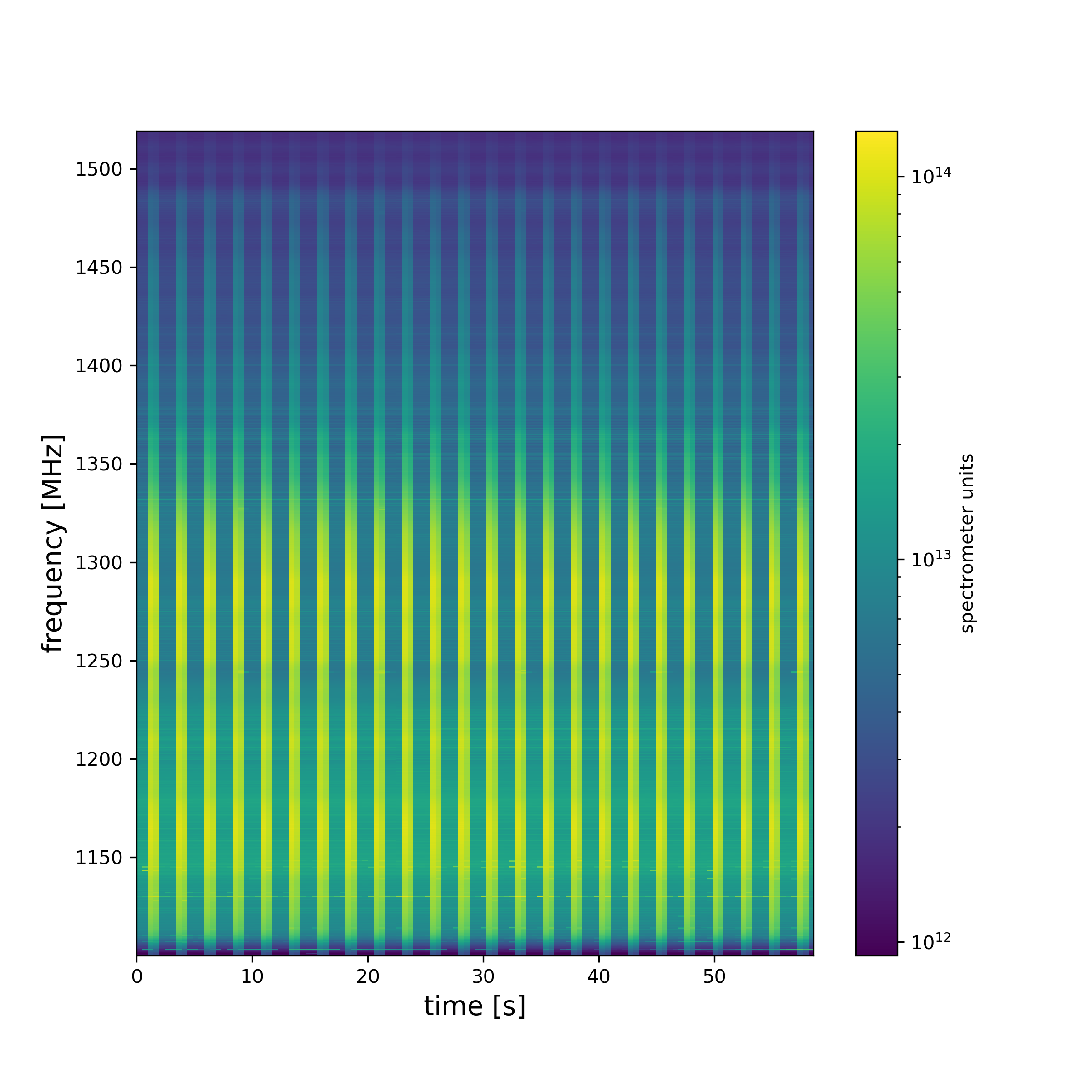} \\
    \includegraphics[height=0.44\linewidth,trim=50 30 160 30, clip]{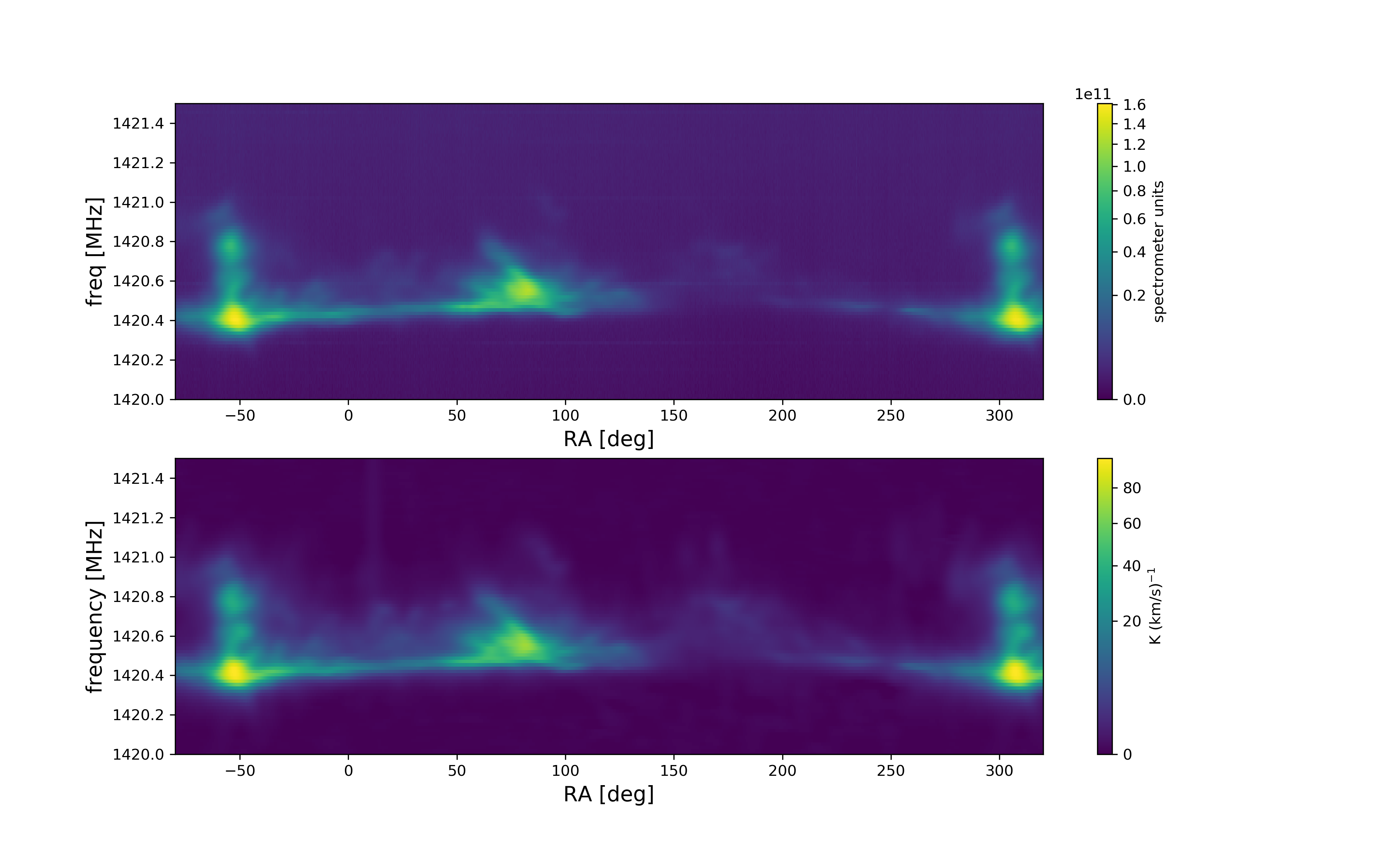} &
    \includegraphics[height=0.40\linewidth,trim= 0 30 0 50, clip]{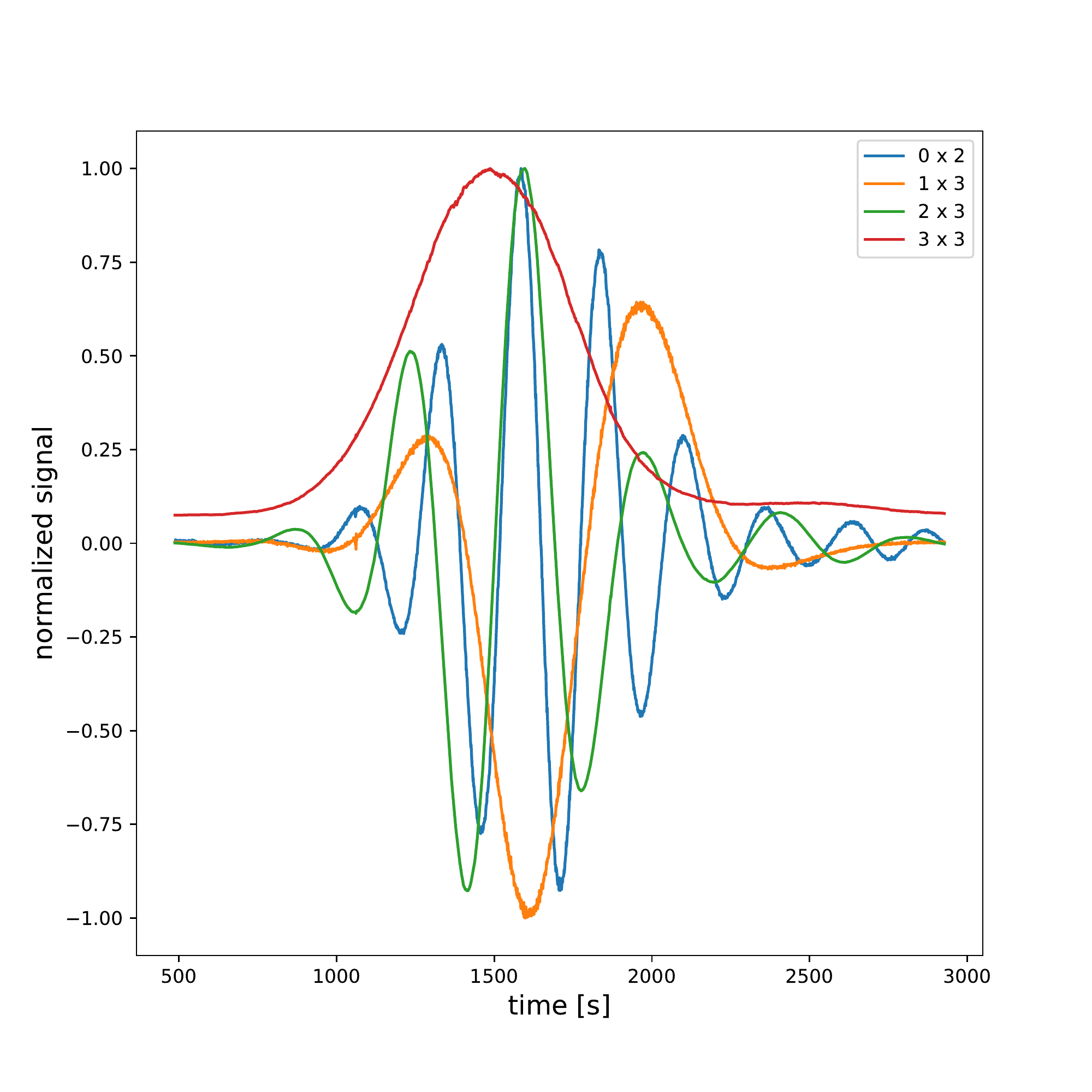} \\

  \end{tabular}
  \caption{BMX testbed prototype telescope at BNL designed to aid
    laboratory test as we perform R\&D in preparation for PUMA: Upper
    figure: Configuration of the system composed of four off-axis
    dishes forming a transit radio interferometer operating at 1.1-1.5
    GHz.  Middle-left: Launch of a fixed-wing drone prototype for
    primary beam calibration; Middle right: Time switching of the beam
    calibration system payload put in the view of one of the BMX
    dishes; Lower left: Galactic 21\,cm signal as seen by BMX (top)
    and predicted based on HI4PI dataset \cite{1610.06175}; Bottom
    right: fringes observed as navigational satellite passes overhead
    for a few selected baselines (all baselines are operational).  }
  \label{fig:bmx}
\end{figure}

\begin{figure}
    \centering
    \includegraphics[width=0.95\textwidth]{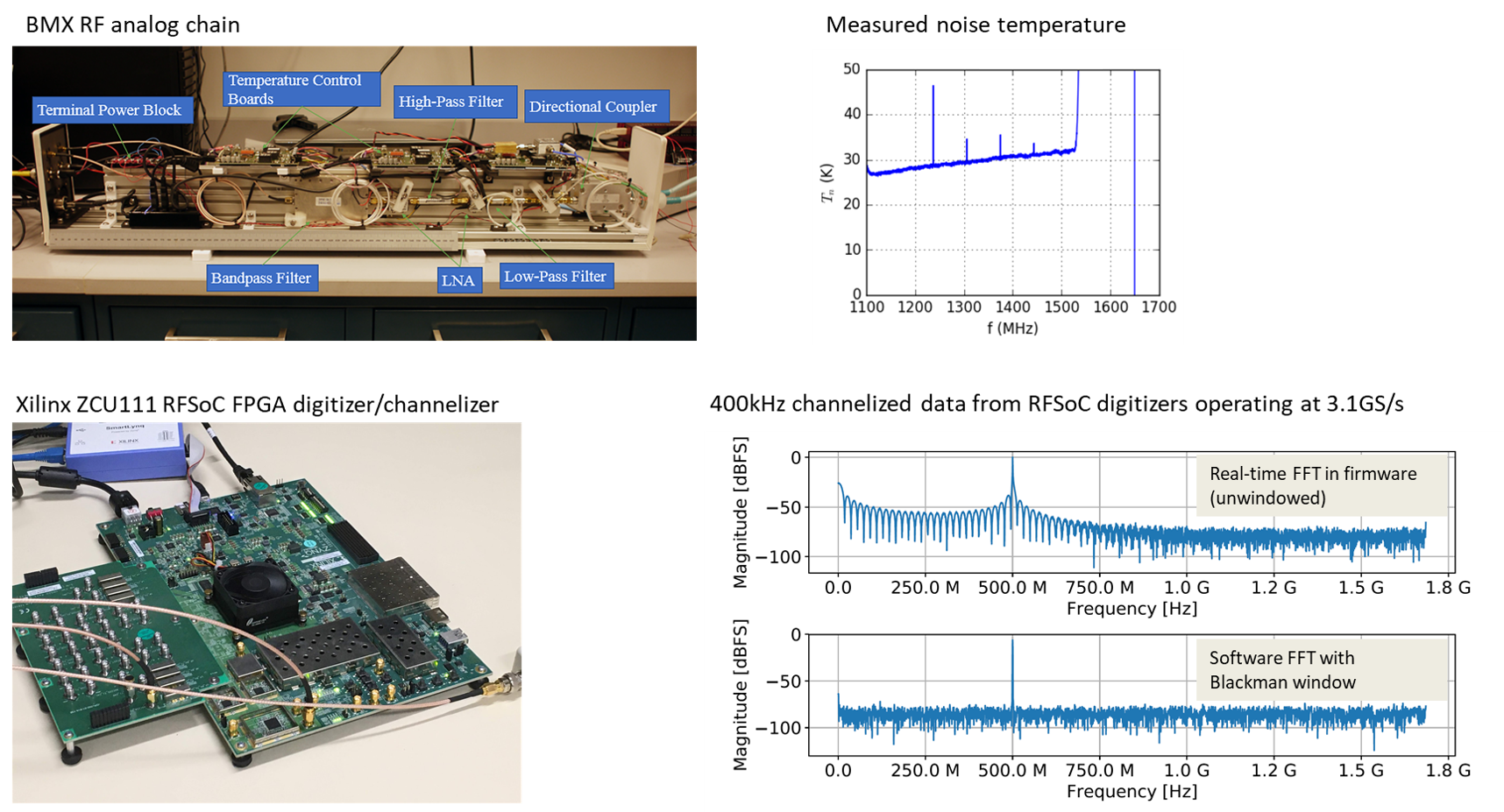}
    \caption{Upper left: analog signal chain for the BMX dual-polarization receivers. Upper right: measured noise performance in lab. Lower left: ZCU111 evaluation board with Xilinx Ultrascale+ RFSoC. Lower right: RFSoC digitized data at 3.1\,giga-sample/s, channelized in firmware in real time (upper panel), channelized in software with Blackman window (lower panel). Firmware development to include lowpass filtering, windowed/polyphase FFT, decimation, and serialization in progress.}
    \label{fig:BMX_elx}
\end{figure}

Within the DOE laboratories, two exploratory R\&D activities have been funded.
\begin{itemize}
    \item Fermilab is a member of the Tianlai collaboration and has contributed to the analysis of correlator data.
    \item At Brookhaven National Laboratory, in addition to supporting the PUMA concept development, a small 4-dish test bed interferometer has been built (BMX). It is being used to develop custom integrated front-end prototypes and to study beam calibration with UAVs. See Figures \ref{fig:bmx} and \ref{fig:BMX_elx}.
\end{itemize}

\subsection{PUMA proto collaboration}

Following the successful CMB-S4 model, we intend to establish a proto-collaboration in the style of HEP collaborations, that will eventually organically grow into the full collaboration as the project transitions from R\&D towards actual project development.

The proposed organizational chart of this entity is outlined in Fig.~\ref{fig:orgchart}. In developed projects, there is a clear distinction between project and science collaboration -- we believe that it is premature for PUMA to adopt this now. Instead, the work will be organized around Pre-Project Topical Teams that will develop individual subsystems as research. The Pre-project Steering committee will be composed of upper management of individual labs and other participating universities and act as champions for the project to funding agencies while ensuring sufficient seed funding. This entire structure could be organized around ``DOE consortia'' concepts before the effort receives project status.  Finally, the Science and Technology Advisory Panel will be composed of members of current projects in the same area and ensure that our R\&D is consistent with development in the field using operational telescopes.

\subsection{Major risks to the development of the facility.}
\label{sec:major-risks-devel}

Programatically, the main risk to the concept would be that the program is rejected by either the Decadal Survey or Snowmass / P5 process. This would not allow DOE to proceed with this program for the foreseeable future, leading to delay of the start of the project for a decade. Mitigation would require a major rethinking of the project and interesting science that can be done in this field in 2030s.

\noindent Technically, we identify three major risks:

\minisection{Array is not calibratable.} Our computer simulations
might show that the system, even if perfectly stable, is simply not
calibratable to the required precision using available information. If
the number of degrees of freedom in the calibration vector is too large,
or contains unexpected degeneracies, it might be impossible to
determine them with sufficient precision. Unavailability of sufficiently
deep catalogs of point sources used for calibration \cite{aguirre2019roadmap}
is also of real concern. \emph{Mitigation Strategy} will depend on the exact nature of the problem. It might involve re-design of certain hardware components (for example addition of outrigger stations for better separation of point sources) or development of new calibration hardware. 

\minisection{Array is not sufficiently stable.} Even if we can perform a calibration at certain point in time, the calibration vector
might drift in time faster than the real-time calibration scheme can correct for it.  \emph{Mitigation Strategy:} Identify the sources of drift and incrementally re-design the relevant components.

\minisection{The FFT correlation might not be feasible in practice.} The majority of up-coming radio interferometers, including the largest proposed ones such as DSA-2000, still rely on direct $N^2$ calibration. At the scale of PUMA, even for the 5K version,  this becomes cost prohibitive. CHIME is currently using real-time calibration and redundant-baseline co-addition, which is equivalent in terms of information content and calibration requirements to FFT correlation. This system is already deployed for transient detection and the work is in progress to demonstrate it for intensity mapping. Experience from CHIME will be folded into our simulations to better understand performance and requirements. \emph{Mitigation Strategy:} Study the science impact of computing intensity lowering strategies: full correlation of a subset of elements, sacrificing baselines that are not crucial for science; lowering the instantaneous bandwidth; lowering the element count and perhaps enabling heterogenenous dish sizes. Using these approaches it might be possible to re-design the system with only a modest impact on the science.

For projects of this scale, funding agency guidance would require a rigorous R\&D program focused on mitigating and retiring major risks before proceeding to construction. We have currently proposed five years of R\&D for PUMA-32K, where the dominant way of retiring the FFT correlation and real-time calibration is through extensive computer simulations. In this way, we will simultaneously retire all three major risks listed above, because a calibratable and sufficiently stable array is a precondition for successful  FFT correlation.

We note that the total system simulation effort at the proposed fidelity has not been performed for any of the Stage-1 low-redshift 21cm experiments and that the EoR community is only now embarking on a similar exercise. For PUMA, extensive simulations in advance of final design are a crucial part of the risk retirement strategy.

By comparing against results and measurements from our lab development (R\&D prong \#1) and smaller hardware prototypes (R\&D prong \#3), these full-system simulations should give confidence in algorithmic correctness and convergence of real-time calibration.  There is already a good understanding of the level of foregrounds and likely incompleteness of current sky catalogs, and we stress that a precise model for the true signal is not needed to validate the approach. Therefore, with a carefully-constructed program of computer simulations we believe we can validate the system design.

After a five-year R\&D phase we will reevaluate its outcomes. This can result in construction, extension of the R\&D phase or de-scope and change of the instrument design.

\section{Conclusions}

PUMA would enable dramatic advances in six basic science drivers that explore three different areas of physics: measuring the expansion history of the Universe and the growth of cosmic structure; probing cosmic inflation through primordial non-Gaussianity and relic inflationary features; discovering fast radio bursts and monitoring pulsars.  The first four drivers are enabled by an intensity mapping survey of neutral Hydrogen out to $z\simeq 6$, which would be the equivalent of a galaxy redshift survey of 2.9 billion galaxies over the same redshift range.  The last two are enabled by PUMAs large field of view and continuous monitoring.  The instrument would also be powerful enough that other exciting science cases, not anticipated here, are highly likely to be developed in the next decade, including time-domain and multi-messenger science.  The required sensitivity and layout of the instrument follows directly from the six science cases, leading to an interferometric, five-year survey employing an array of 32,000 six-meter parabolic dishes located on the sites of a hexagonal close-packed lattice with 50\% occupancy.

The most difficult to achieve goals are those related to intensity mapping.  Hydrogen intensity mapping is a relatively new technique with many advantages that leverages the tremendous advances in computing and wireless communication devices that have occurred over the last decade.  However, while first generation experiments are under construction or underway, there is still a lot of R\&D required before an experiment such as PUMA could be constructed and operated.  Key areas of research include calibration, correlator architectures, RFI mitigation, foreground mitigation, and simulation capability.  We plan a 4-pronged R\&D effort with simultaneous technology development in the laboratory, development of computing simulations, software and pipelines, development of real-time signal processing and calibration and of small path-finder arrays.  This R\&D provides a clear path that will lead to a possible realization of PUMA.  We have outlined a conceptual/reference design that is informed by our science goals and sufficient to enable preliminary costing and trade-off studies.  We expect this design will evolve as our R\&D matures and as we incorporate lessons from the existing $21\,$cm IM facilities.

\newpage

\appendix

\section{Computational Intensity}
\label{app:ccost}

To calculate relative computational intensity of the individual baseline correlator and the FFT correlator, we use a simplified analysis that counts floating point operations. This should only be taken as an order of magnitude calculation, but it does give not just the relative intensity of the different techniques, but also a rough guide to the complexity of the task at hand.

\subsection*{Direct correlation}

Assuming Nyquist sampling the data sample rate is $2B$, where $B=900$\,MHz is the system bandwidth. Each pair of samples needs to be multiplied and added (2 operations) and there are 4 Stokes parameters to calculate. The floating point operations rate is thus given as $16B$ per baseline. For $N_A$ antennae, the total number of antenna pairs is $\sim N_A^2 / 2$ and if we calculate the full correlation for the fraction $f\sim 10^{-2}$ we get the total computation cost as
\begin{equation}
  C = 8 f N_A^2 B
\end{equation}
This results in $\sim 2$ PFLOPS for PUMA-5K and $70$ PFLOPS for PUMA-32K. We assume this as the dominant computational cost for the correlator and that additional processing for the calibration is incremental.

\subsection*{FFT correlator}

The computational intensity of a real FFT of size $N$ is $(5/2) N \log_2 N$ for the radix-2 Cooley–Tukey FFT algorithm. We will use the same equation also for combinations that are not purely radix-2. For a 2D FFT alogorithm on an $N\times M$ block, the total number of operations is  $\propto N \times M\log_2M + M \times N\log_2N = NM \times \log_2 NM = N_L \times \log_2 N_L$, where $N_L$ is the number of lattice sites on which the FFT is performed.

To correctly estimate the computing requirements for the FFT correlator, we need to take the following factors into account when calculating the appropriate $N_L$:

\begin{itemize}
\item The FFT based correlation algorithm automatically correlates all lattice sites, whether they are occupied or not

\item A hexagonal closely-packed lattice maps to a 50\% filled square-packed lattice after an anisotropic stretch.

\item In our forecast we assumed the array shape to approximate a circle, whereas the FFT area needs to fit inside a rectangle.

\item To avoid coadding baselines which appear short in the periodic FFT grid, we need to pad the FFT size by a factor of 2 in each dimension \cite{1710.08591}.

\item We further need to round the FFT sizes to integer numbers of the form $2^i3^j$ for efficient FFTs (although factors of 5 and 7 might also be permissible).
\end{itemize}
The first 4 factors demand FFTs with approximately $32/\pi/({\rm fill\ factor})\sim 20$ more lattice sites than there are antennae, while the last one increases this by another $\sim 10\%$. We round this to number of lattice sites being equal to $N_L = 20N_A$.

 To get the total computational cost, we just count the total rate of signal samples (degrees of freedom). For two polarizations, this is $4B N_A$, so we need to process FFTs at the rate of $4B$ with the computational cost of $(5/2) N_L \log_2 N_L$ each. With $N_L\sim 20N_A$ we have
\begin{equation}
  C = 200 B\ N_A \log_2 \left( 20 N_A \right)
\end{equation}
or $\sim 15$ PFLOPS for PUMA-5K and $\sim 100$ PFLOPS for PUMA-32K.

This is a conservative estimate, taking into account all the inefficiencies. Assuming more efficient algorithms can be used for FFTs on hexagonal lattices and with an overall array shape that is a better match, the prefactor to $N_A$ could drop from $\sim 20$ to $\sim 8$.
We assume the FFTs to be the dominant computing cost for the FFT correlator and that the processing of output FFT grids  will add only incrementally to the total computational intensity.

\section{Data rates}
\label{app:drates}

To calculate the data rates, let's assume we output all coadded baselines at the rate at which they changes substantially.

For the E-W baseline on length $d$, the phase difference changes as $d \cos\alpha (\dot{\alpha} / \lambda)$, where $\alpha$ is the angle with respect to zenith. Nyquist sampling would require us to sample changes every $\pi$ radians, but let us assume we want a safety factor of $\times 2$ for systematics control. The phase changes the fastest at zenith ($\alpha=0$), where it moves by  $\pi/2$ every $\delta T = \lambda / (4 d \dot{\alpha})$. In the worst case scenario, i.e.\ at 1100MHz, pointing at the celestial equator with the longest baseline, we find $\Delta T \simeq 1.6$s for PUMA-5K and $\Delta T \simeq 0.6$s for PUMA-32K. We will therefore assume a correltor output every $\Delta T\simeq 1$ second.

Each $u-v$ point is described by 4 numbers (for each Stoke's parameter, each taking 8 bytes), so the general equation for data rates is
\begin{equation}
  R \sim 32 N_{\rm base} N_{\rm ch} / \Delta T,
\end{equation}
where $N_{\rm base}$ is the number of baselines we want to save.
For the direct correlator, $N_{\rm base}=N_A^2/2$, so we have 
\begin{equation}
  R \sim 16 f N_A^2 N_{\rm ch} / \Delta T.
\end{equation}
Since the main purpose of direct correlation is only to assist with real-time calibration, we will likely save just a fraction of the data for quality assurance purposes. This might require special schemes, such as saving fewer baslines, but saving them more often.

The number of redundant baselines is given (to leading order) by $N_{\rm base}= 2N_A/(\rm fill\ factor) = 4N_A$ for PUMA. To see this consider at all possible baselines an antenna at the edge of an array can make and multiply by two for mirroring in one axis (mirroring in both axes results in the same baseline up to the sign). We have checked this result with an explicit calculation. We therefore have
\begin{equation}
  R \sim 128 N_A \times N_{\rm ch} / \Delta T.
\end{equation}

For $N_{\rm ch}=2\times 10^4$ (giving velocity resolution of $\sim 70$km/s at the longest wavelengths), we have $R\sim 13 $\,GB/s for PUMA-5K and $\sim 82$GB/s for PUMA-32K. This amounts to $1$ PB/day for PUMA-5K and $7$PB/day for PUMA-32K. This is a conservative estimate -- by averaging spectral bins at higher frequency, saving long baselines less often and other techniques, we can save a factor of a few. Therefore we give this as the upper range on the total data-rate.

\section{Cost Model, Basis of Estimate, Summary, and Timeline}
\label{app:cost}
Although the design of PUMA is at the pre-conceptual level, we may make reasonable, conservative assumptions about scaling and future technology developments to extrapolate from precursor dish array projects and arrive at plausible bounds on the eventual project cost.  As the design matures and the project is informed by R\&D and experience of precursor \stageone\ surveys, a more thorough and parametric bottom-up cost estimate will be developed. 
For the purposes of the Astro 2020 review, we made a first-pass estimate based on a parametric costing model. This model relies, where possible, on available component costs (primarily computing, electronic hardware and dish construction) using data supplied by the current generation of experiments, and otherwise on extrapolations based on projects of comparable scope that the authors participate in. Where engineering judgement was required, an attempt was made to be conservative.  It is our intention to update and document the evolution of the cost model as the proposed project matures; the latest model can always be found at \url{http://puma.bnl.gov/Costing}.

We costed PUMA-5K and PUMA-32K as independent projects, although the smaller program could be built as a first stage of the full array. 
Hardware costs used available data from prior intensity mapping experiments where possible. For key microelectronics components, we extrapolated industry published cost trends over the past 5 - 15 years to the proposed procurement dates. Labor costs were estimated by extrapolating from prior and proposed projects of similar scale (e.g.\ DESI, HERA-II, CMB-S4, and LSST). 
As of this point estimate, construction and commissioning costs were estimated at \$59\,mil and \$373\,mil in FY19 dollars for PUMA-5K and PUMA-32K, respectively.

\subsection{Cost Estimate Detail}
We assume an aggressive R\&D program lasting 4 or 5 years, addressing the most critical analysis issues and bringing key technologies to maturity as discussed in Sections 4 - 6. At the conclusion of this phase, the project would be ready for site-specific agency reviews (DOE-CD1, NSF-PDR). Following R\&D, we envision a 2-year Final Design phase including negotiations with the host country over site acquisition.  We project the construction and commissioning of PUMA-5K to require 4 years. A subsequent two-year build-out of PUMA-32K could occur without completely disrupting the operation of PUMA-5K, for instance by observing at night only.

Key elements of the array construction cost estimate are: an existing radio-quiet site with road, power, and fiberoptic communication access; non-tracking, altitude-adjustable $6\,$m molded fiberglass dish array with on-antenna front end dual-polarization RF amplifier and filter chains, on-antenna digitizers operating in the first Nyquist zone with F-engines implemented in FPGA, and fiberoptic serial IO. For these components we extrapolated from actual costs from HIRAX, CHIME, and BMX. Industry pricing data was used to apply 5\%/year and 10\%/year cost trends for the digitizer and FPGA/F-engine/SERDES, respectively.
Per-link cost was adopted from SKA-Mid estimate for a sub-picosecond optical synchronization link. Such timing systems are expected to be needed for many upcoming research and industrial applications; we assume 8\%/year cost reduction factor. In PUMA, precision timing links will be installed to distribution boxes each serving a cluster of 6 antenna stations in close proximity. We estimate that this will be sufficient to maintain phase coherence over the entire array. If RFI considerations make it impractical to locate the digitizers and FPGAs on each antenna, they could alternatively be housed in the well-shielded per-cluster enclosures used for timing fanout.

The computational cost of a conventional FX correlator scales as
    $$R = 2B(N \log_2 N_\mathrm{ch} + N^2)$$
Where $R$ represents the aggregate multiplication rate, $B$ the bandwidth, $N_{\rm ch}$ the number of frequency channels processed, and $N$ the number of antennae \cite{warnick2018phased}. The first and second terms correspond to the F- -and X-engines respectively. For large-$N$ intensity mapping instruments, the X-engine term is dominant and $R \sim 2BN^2$. To verify this scaling model, we used data from CHIME/HIRAX \cite{LBNPrivate} and DSA-2000 \cite{hallinan2019dsa}.  As discussed in Section \ref{subsec:FFT_correlator},  a full $N^2$ correlator will be impractical from a cost and power standpoint even at the scale of PUMA-5K. Even with the most optimistic assumptions about technology advances PUMA-32K will need a direct-imaging, FFT correlator back end. The FFT correlator's computational cost will scale roughly as $N_s\log_2N_s$, where $N_s$ is the number of lattice sites taking into account padding, gridding, etc.\ ($N_s\sim 20N$, see Appendix \ref{app:ccost}). To account for periodic gain and phase calibrations, we assume the correlator will also perform full $N^2$ correlation on  1\% of the baselines, resulting in a correlator cost of that scales as $(N_s\log_2N_s + 0.01N^2)$. A cost deflator of 5\%/year was applied to this element. Cost for the additional correlator subsystems (FRB/pulsar backends) were estimated at 10\% of the main correlator.

Costs for control, calibration, data management, installation, commissioning, operations, and management were taken as a percentage of total project construction cost using available data from comparable large agency-funded experiments; further details are present in the costing documents linked above.

\bibliographystyle{utphys}
\bibliography{references}
\end{document}